\documentclass[%
 reprint,
superscriptaddress,
bibnotes,
amsmath,amssymb,
]{revtex4-2}

\usepackage[utf8]{inputenc}
\usepackage{siunitx}
\usepackage[english]{babel}			
\usepackage{graphicx}				
\usepackage{amssymb,amsmath}		
\usepackage{float}                  
\usepackage{braket}
\usepackage{bm}
\usepackage{hyperref}
\hypersetup{colorlinks=true,allcolors=blue}
\usepackage{lineno}

\newcommand{\inext}{_\text{ext}}

\AtBeginDocument{\RenewCommandCopy\qty\SI}

\preprint{APS/123-QED}

\begin{document}

\date{\today}

\title{Entanglement-induced collective many-body interference}

\author{Tommaso Faleo}
\email{Tommaso.Faleo@uibk.ac.at}
\affiliation{Institut für Experimentalphysik, Universität Innsbruck, Technikerstr. 25, 6020 Innsbruck, Austria}
\author{Eric Brunner}
\altaffiliation{Current address: Quantinuum, Partnership House, Carlisle Place, London SW1P 1BX, United Kingdom}
\affiliation{Physikalisches Institut, Albert-Ludwigs-Universit{\"a}t Freiburg, Hermann-Herder-Straße 3, 79104 Freiburg, Germany}
\author{Jonathan W. Webb}
\affiliation{Institute of Photonics and Quantum Sciences, School of Engineering and Physical Sciences, Heriot-Watt University, Edinburgh, EH14 4AS, UK}
\author{Alexander Pickston}
\affiliation{Institute of Photonics and Quantum Sciences, School of Engineering and Physical Sciences, Heriot-Watt University, Edinburgh, EH14 4AS, UK}
\author{Joseph Ho}
\affiliation{Institute of Photonics and Quantum Sciences, School of Engineering and Physical Sciences, Heriot-Watt University, Edinburgh, EH14 4AS, UK}
\author{Gregor Weihs}
\affiliation{Institut für Experimentalphysik, Universität Innsbruck, Technikerstr. 25, 6020 Innsbruck, Austria}
\author{Andreas Buchleitner}
\affiliation{Physikalisches Institut, Albert-Ludwigs-Universit{\"a}t Freiburg, Hermann-Herder-Straße 3, 79104 Freiburg, Germany}
\affiliation{EUCOR Centre for Quantum Science and Quantum Computing, Albert-Ludwigs-Universität Freiburg, Hermann-Herder-Straße 3, 79104 Freiburg, Germany}
\author{Christoph Dittel}
\affiliation{Physikalisches Institut, Albert-Ludwigs-Universit{\"a}t Freiburg, Hermann-Herder-Straße 3, 79104 Freiburg, Germany}
\affiliation{EUCOR Centre for Quantum Science and Quantum Computing, Albert-Ludwigs-Universität Freiburg, Hermann-Herder-Straße 3, 79104 Freiburg, Germany}
\affiliation{Freiburg Institute for Advanced Studies, Albert-Ludwigs-Universität Freiburg, Albertstraße 19, 79104 Freiburg, Germany}
\author{Gabriel Dufour}
\affiliation{Physikalisches Institut, Albert-Ludwigs-Universit{\"a}t Freiburg, Hermann-Herder-Straße 3, 79104 Freiburg, Germany}
\author{Alessandro Fedrizzi}
\affiliation{Institute of Photonics and Quantum Sciences, School of Engineering and Physical Sciences, Heriot-Watt University, Edinburgh, EH14 4AS, UK}
\author{Robert Keil}
\email{Robert.Keil@uibk.ac.at}
\affiliation{Institut für Experimentalphysik, Universität Innsbruck, Technikerstr. 25, 6020 Innsbruck, Austria}
\begin{abstract}

Entanglement and interference are both hallmark effects of quantum physics. Particularly rich dynamics arise when multiple (at least partially) indistinguishable particles are subjected to either of these phenomena. By combining both entanglement and many-particle interference, we propose an interferometric setting through which $N$-particle interference can be observed, while any interference of lower orders is strictly suppressed.
We experimentally demonstrate this effect in a four-photon interferometer, where the interference is nonlocal, in principle, as only pairs of photons interfere at two separate and independent beam splitters. A joint detection of all four photons identifies a high-visibility interference pattern varying as a function of their collective four-particle phase, a genuine four-body property.

\end{abstract}

\maketitle

\section*{introduction}

Entanglement is arguably one of the most fascinating and powerful phenomena arising in quantum physics. If two or more quantum objects are entangled they can no longer be described as independent entities, instead, their properties are interlinked through their associated degrees of freedom (dof), regardless of how these dof are measured and how far apart the objects are from each other \cite{Horodecki2009}. This effect has been famously and unambiguously demonstrated in Bell inequality tests \cite{Bell1964,Hensen2015} and it enables a rich variety of applications within the fields of quantum communication \cite{Briegel1998,Chen2008,Nadlinger2022}, quantum computation \cite{Knill2001,Raussendorf2003,Bartolucci2023} and quantum simulation \cite{Sansoni2012,Matthews2013}. 

A similarly fundamental effect is given by two-particle (2P) interference, as first observed by Hong, Ou, and Mandel (HOM) via the absence of photon pair coincidences at the output of a balanced beam splitter \cite{HOM1987}. HOM interference has emerged as one of the most widely employed quantum effects in the field of photonics \cite{Bouchard2021} and it requires indistinguishability of the involved particles.
Specifically, the particles should not be identifiable by their internal states --- the states of those dof which do not participate in the dynamics --- thus forbidding the retrieval of 2P which-way information \cite{Tichy2014, Dittel2021}.
The HOM experiment represents a starting point for the exploration of multi-particle interference phenomena, and its generalisation to larger-scale $N$-particle ($N$P) systems \cite{Lim2005,Tichy2010,Tichy2011} finds applications in fundamental tests of quantum mechanics \cite{Pleinert2021} and in the advancement of quantum technologies \cite{Flamini2019,Madsen2022}.
However, $N$P interference for $N>2$ is inherently more complex than the HOM effect.
The rich spectrum of multi-particle interference terms stemming from the various particle-exchange processes gives rise to nontrivial behaviours, especially when partial distinguishability among the particles is introduced \cite{Tichy2011,Tichy2013,Young-SikRa2013}.
Therefore, a substantial challenge is to discern genuine $N$P interference, which has, in particular, been tackled by examining specific signatures of this interference \cite{Walschaers2016,Giordani2018,Pont2022,brunner_many-body_2022,Seron2023} and identifying instances of totally destructive $N$P interference which generalise the HOM scenario \cite{Tichy2010,Crespi2015,Dittel2017,Dittel2018,Muenzberg2021}.
Recent experimental observations \cite{Menssen2017,Jones2020} have, furthermore, demonstrated the presence of genuine multi-particle interference witnessed by its dependence on a collective geometric phase that is determined by the overall internal quantum state of all of the particles.
The emergence of this collective $N$P phase can be attributed to carefully engineered internal states of the particles as well as the specific structure of the interference terms, involving cyclic permutations of all $N$ particles \cite{Shchesnovich2018}.  

Here, we propose and experimentally demonstrate a novel interference effect to realize genuine $N$P interference by combining the concepts of entanglement and many-body interference. In particular, we show that by using entanglement between a subset of particles at the input of separate interferometers, full contrast interference fringes emerge for the $N$P correlator as a function of a collective $N$P phase, while the signal of all lower-order correlators remain independent of the involved phases. In contrast to the realisation of $N$P phases of previous approaches \cite{Menssen2017,Shchesnovich2018,Jones2020}, which necessitate the engineering of an $N$-dimensional internal Hilbert space and the mixing of all particles in an $N$-cyclic interferometer, here only a two-dimensional internal Hilbert space and pairwise exchanges of particles at separate beam splitters are needed to induce collective $N$P interference. By virtue of its design, the entanglement-induced collective $N$P interference effect exhibits full interference contrast for arbitrarily large numbers of particles.

The genuine $N$P interference effect presented here is by construction independent of the spatial separation between the beam splitters, such that the resulting collective many-body interference can be realised in a nonlocal fashion.
Multi-particle interference in disjoint interferometers is otherwise only known from Franson-interference \cite{Rice1994,Agne2016}, which relies on maximally entangled states in the energy-time dof intrinsically linked to the photon generation process.
In contrast, the entanglement-induced collective phase shown in this work can be imprinted on any bosonic state with a two-dimensional internal Hilbert space.

\section{theory} \label{section:theory}

\begin{figure*}
\centering
\includegraphics[width=\linewidth]{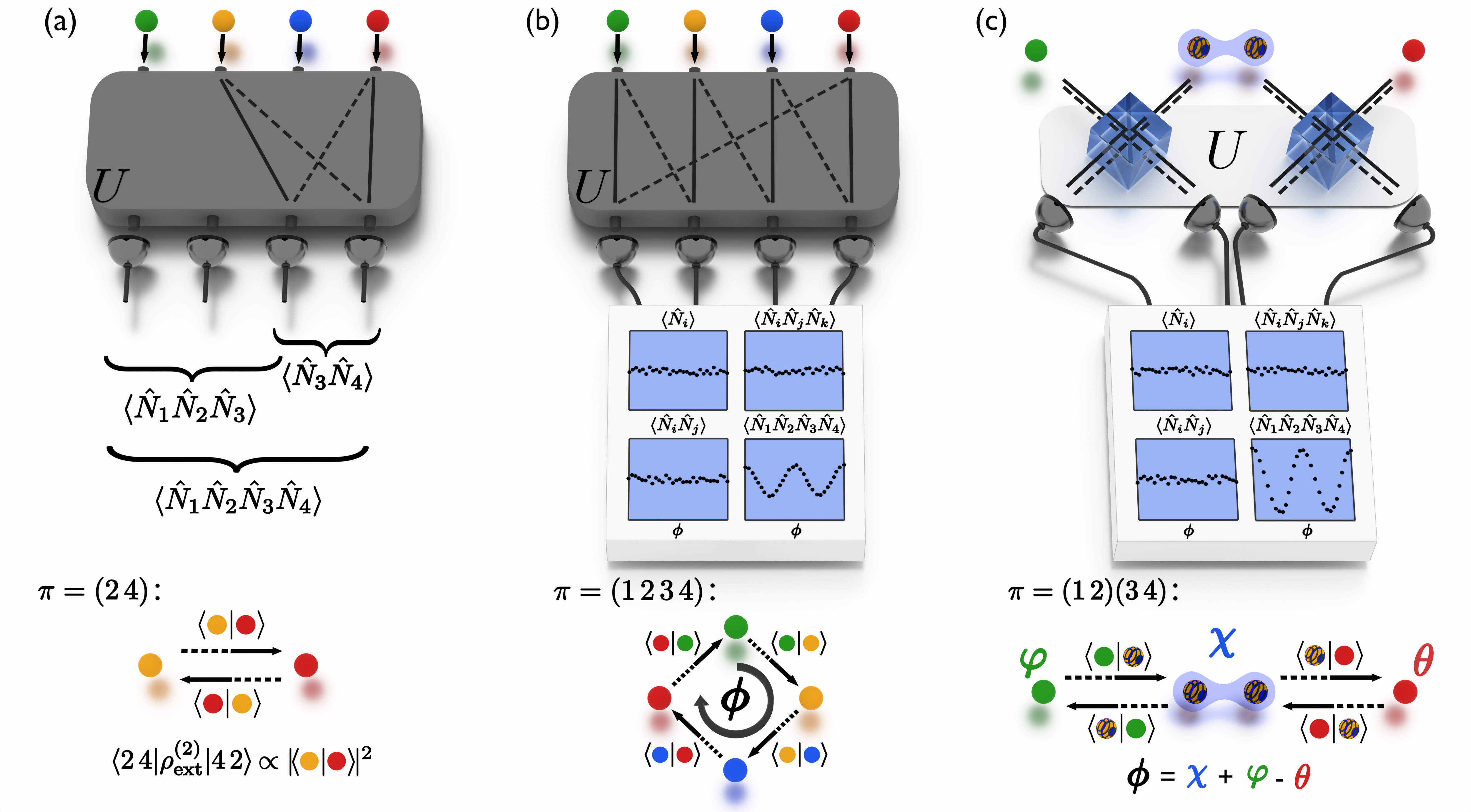}
\caption{Many-particle interference and collective phases.
\textbf{(a)} In a fully connected interferometer with single-particle unitary matrix $U$ (here 4x4), multiple exchange processes among combinations of input-output channels contribute to specific output events.
The reduced 2P state associated with the second (yellow) and fourth (red) particles contributes to the two-point correlator $\braket{\hat{N}_3\hat{N}_4}$ via the $2$P transitions indicated by the solid and dashed lines. The interference of these two $2$P paths is encoded in the $2$P coherence at the bottom of the panel, which is associated with the permutation $\pi=(2\,4)$ obtained by taking one $2$P path in the forward direction followed by the other one in the backward direction \cite{Tichy2015}.
The matrix elements of $U$ weight these contributions, as well as the contributions from all other 2P sets. Similar considerations apply to all correlators from combinations of different output modes.
\textbf{(b)} The four-particle (4P) coherence corresponding to the depicted 4-cyclic permutation leads to a genuine 4P interference that depends on a collective phase $\phi$. This collective phase is set by the summation of phases $\phi_{ij}$ (see main text) resulting from the overlaps of the particles' internal states along a ``circle-dance'' graph representative of the permutation process \cite{Jones2020,Shchesnovich2018}, as depicted at the bottom of the panel.
\textbf{(c)} A genuine entanglement-induced 4P interference can be achieved by interfering entangled particles (blue envelope) with particles in separable states in two independent and separate beam splitters. In this process, the entanglement induces a collective 4P phase term $\phi$, set by the internal states of all particles, through the two 2P permutations of $\pi=(1\,2)(3\,4)$. The collective phase $\phi$ only affects the four-point correlator and introduces full contrast interference fringes: $\braket{\hat{N}_1 \hat{N}_2 \hat{N}_3 \hat{N}_4} \propto \cos^2 \left( \phi/2 \right)$ (see main text).}
\label{fig:GenericScheme}
\end{figure*}


\noindent \textbf{Many-body interference and collective phases}\\
In a general interference scenario among $N$ particles, as depicted in Fig.~\ref{fig:GenericScheme}(a) for $N=4$ unentangled particles, each input mode $m_i=1 \dots N$ of the multiport interferometer is occupied by a single particle, which carries some internal dof (illustrated by the particle's colour in Fig.~\ref{fig:GenericScheme}).
Interference effects across the multiport manifest through correlations of the particle number at the output of the interferometer.
We specifically consider correlations between the occupations of $k \leq N$ distinct modes $p_1, \dots, p_k$. 
Denoting by $U$ the single-particle unitary matrix describing the transformation between input and output modes (henceforth termed external modes) implemented by the interferometer, the expectation value of such a $k$-point correlator is given by
\begin{equation}\label{eq:kpointcorrelator}
\braket{\hat{N}_{p_1} \dots \hat{N}_{p_k}} = \sum_{\bm{m},\bm{n}} \prod_{i=1}^k U_{p_i m_i} U_{p_i n_i}^\ast \braket{\bm{m} | \rho\inext^{(k)} | \bm{n} } \,,
\end{equation}
where the external number operator $\hat{N}_p$ counts the number of particles in output mode $p$ irrespective of their internal state. In terms of the operator $\hat{a}_{p\alpha}$ annihilating a particle with internal state $\ket{\alpha}$ in external mode $p$, the external number operator reads $\hat{N}_p = \sum_\alpha \hat{a}_{p\alpha}^{\dag}\hat{a}_{p\alpha}$, where the sum runs over a basis of the internal Hilbert space.

Since it involves products of $k$ creation and $k$ annihilation operators, the $k$-point correlator $\hat{N}_{p_1} \dots \hat{N}_{p_k}$ is an example of a $k$-particle ($k$P) observable \cite{brunner_signatures_2018,dufour_many-body_2020,brunner_many-body_2022}. As a consequence, for $k<N$, it only probes the reduced state $\rho\inext^{(k)}$, obtained by tracing out the internal dof and $(N-k)$ particles from the input state \cite{brunner_many-body_2019,Minke2021,brunner_many-body_2022}.
Specifically, the matrix elements $\braket{\bm{m} | \rho\inext^{(k)} | \bm{n} }$ appear in Eq.~\eqref{eq:kpointcorrelator}, with $\ket{\bm{m}} = \ket{m_1} \otimes \dots \otimes \ket{m_k}$ the tensor product of the external single-particle basis states $\ket{m_i}$, corresponding to the particle occupying input port $m_i$ of the interferometer, and analogously for $\ket{\bm{n}}$.
The nonzero matrix elements $\braket{\bm{m} | \rho\inext^{(k)} | \bm{n} }$ are indexed by tuples $\bm{m}$ and $\bm{n}$ that are connected by a permutation $\pi\in\mathrm{S}_k$ in the symmetric group $\mathrm{S}_k$ of their $k$ entries: $\bm{n} = \pi (\bm{m}) = (m_{\pi^{-1}(1)}, \dots, m_{\pi^{-1}(k)})$.
The off-diagonal matrix elements ($\bm{m}\neq\bm{n}$) are called \textit{$k$P coherences}.

In the traditional scenario where each particle in mode $m_i$ is associated with a particular internal quantum state $\ket{\xi_{m_i}}$, corresponding to an input product state $\ket{\Psi}=\hat{a}_{m_1,\xi_{m_{1}}}^{\dag} \dots \hat{a}_{m_N,\xi_{m_{N}}}^{\dag} \ket{0}$, all matrix elements are set, up to a multiplicative constant, by the overlaps between internal states of particles in the specific $k$ modes:
\begin{equation} \label{eq:overlaps}
    \braket{\bm{m}|\rho\inext^{(k)}|\bm{n}} \propto \prod_{i=1}^k\braket{\xi_{m_i}|\xi_{m_{\pi^{-1}(i)}}}.
\end{equation}
Therefore, these overlaps, i.e., the particles' mutual (in)distinguishabilities, determine all matrix elements.
In general, the overlaps are complex numbers $\braket{\xi_i|\xi_j}=r_{ij}e^{i\phi_{ij}}$, with real amplitude $r_{ij}$ and phase $\phi_{ij}$.
In the two-point correlator $\braket{\hat{N}_3 \hat{N}_4}$ shown in Fig.~\ref{fig:GenericScheme}(a), for example, the indicated transition processes lead to HOM-type interference involving the diagonal term $\braket{2,4| \, \rho\inext^{(2)} \, |2,4} \propto \braket{\xi_{2}|\xi_{2}}\braket{\xi_{4}|\xi_{4}} = 1$ and the $2$P coherence term $\braket{2,4| \, \rho\inext^{(2)} \, |4,2}\propto \braket{\xi_{2}|\xi_{4}}\braket{\xi_{4}|\xi_{2}}=r_{2,4}^2$, corresponding to the permutations (in cycle notation \cite{Rotman1995}) $\pi=(2)(4)$ and $\pi=(2\,4)$, respectively (see bottom of Fig.~\ref{fig:GenericScheme}(a)).
As in HOM (i.e. 2P) interference, no phase dependence arises and only the real amplitude $r_{2,4}^2$, which expresses the particles' pairwise indistinguishability, determines $\braket{\hat{N}_3 \hat{N}_4}$ \cite{HOM1987}.

In more complex multi-particle exchanges, the engineering of internal states $\ket{\xi_i}$ via an $N$-dimensional Hilbert space allows to witness a phase dependence that is set by the specific permutation $\pi$ and the overlaps among particle pairs \cite{Menssen2017,Jones2020}.
Notably, the genuine $N$P interference introduced in \cite{Shchesnovich2018} is encoded in a $N$P coherence $\braket{\bm{m}|\rho\inext^{(N)}|\pi(\bm{m})}$ associated with a $N$-cyclic permutation $\pi=(1\,2\,3 \dots N)$, as shown in Fig.~\ref{fig:GenericScheme}(b), which depends on the “circle-dance” collective $N$P phase $\phi=\phi_{1,2}+\phi_{2,3}+\dots+\phi_{N,1}$.
Engineering zero overlaps of the internal states of non-neighbouring particles in the permutation --- particles in disconnected vertices of the permutation graph --- ensures that no other phase dependence appears in the signal. This, however, requires in  the traditional scenario the control over an (at least) $N$-dimensional internal Hilbert space\\


\noindent \textbf{Scheme for entanglement-induced collective interference}\\
We introduce here a novel scheme, depicted for $N=4$ in Fig.~\ref{fig:GenericScheme}(c), to witness a collective $N$P phase using entanglement.
In this scheme, the interferometer consists of $N/2$ separate beam splitters, where the first one mixes modes 1 and 2, the second one modes 3 and 4, and so on.
Pairwise 2P entanglement in one of the internal dof (here polarisation) between particles in mode 2 and 3, 4 and 5, ..., $N-2$ and $N-1$ can then be used to induce collective interference across the separate beam splitters.
For simplicity, we now restrict our discussion to $N=4$ (as realised in our experiment), while the general case of $N$ particles follows analogously. As input, we consider the combination of a polarisation-entangled photon pair in modes 2 and 3 and an unentangled photon pair in the remaining modes 1 and 4 (see Fig.~\ref{fig:GenericScheme}(c)). This is described by the four-particle ($4$P) state
\begin{equation} \label{eq:inputState}
\ket{\psi} = \frac{1}{\sqrt{2}} \hat{a}_{1,\mathrm{S}}^{\dag}(\varphi) ( \hat{a}_{2,H}^{\dag} \hat{a}_{3,V}^{\dag} + e^{-i\chi} \hat{a}_{2,V}^{\dag} \hat{a}_{3,H}^{\dag}) \hat{a}_{4,\mathrm{S}}^{\dag}(\theta) \ket{0},
\end{equation}
where $\hat{a}^\dagger_{i,H(V)}$ creates a photon in external mode $i$ with horizontal (vertical) polarisation internal state, and
\begin{equation} \label{eq:separable state}
\hat{a}^\dagger_{i,\mathrm{S}}(\varphi) = \frac{1}{\sqrt{2}} (\hat{a}^\dagger_{i,H} + e^{i\varphi} \hat{a}^\dagger_{i, V} )
\end{equation}
sets the polarisation to a balanced superposition in the H/V-basis with phase $\varphi$, and analogously for $\hat{a}^\dagger_{i,S}(\theta)$. 
Note that while Eq.~\eqref{eq:overlaps} no longer holds for entangled input states, the expression of the $k$-point correlator in Eq.~\eqref{eq:kpointcorrelator} remains valid independently of the exact state associated with $\rho_\mathrm{ext}^{(k)}$.

Due to the specific topology of the interferometer in Fig.~\ref{fig:GenericScheme}(c) (defining the single-particle unitary $U$), not all four modes mix, such that many terms in Eq.~\eqref{eq:kpointcorrelator} vanish.
The only $k$P coherences that contribute are associated with permutations $\pi=(1\,2)$, which exchanges modes 1 and 2, $\pi= (3\,4)$, which exchanges modes 3 and 4, and their product $\pi=(1\,2)(3\,4)$.
This factorisation reflects the structure of the interferometer, with two separate beam splitters, which can, in principle, be arbitrarily far apart.
If we assume the input to be given only by the first summation term in Eq.~\eqref{eq:inputState}, the contributions arising from particle exchanges according to the above three permutations are given by $|\braket{\varphi | H}|^2$, $|\braket{V | \theta}|^2$ and $|\braket{\varphi | H}|^2|\braket{V | \theta}|^2$, respectively (see also Eqs.~\eqref{eq:app1_helper1} and \eqref{eq:app1_helper2} in App.~\ref{app:kPcoherences} for a more detailed picture) --- i.e., by the modulus square of the inner products of the exchanged particles' internal states. These are independent of $\varphi, \theta$ and $\chi$.
Analogously, if only the second term of Eq.~\eqref{eq:inputState} was considered, no phase contributions would be present in the four-point correlator~\eqref{eq:kpointcorrelator}.
The phase dependence results from the superposition of both $4$P states.
The additional cross terms are the projection of the first term in~\eqref{eq:inputState} onto the second term exchanged according to $\pi = (1\,2)(3\,4)$ --- i.e. the matrix element $\braket{ 1,2,3,4 | \, \rho^{(4)}\inext \, | 2,1,4,3}$ in Eq.~\eqref{eq:kpointcorrelator} --- and its complex conjugate (c.c.).
This matrix element is proportional to $e^{-i\chi} \braket{\varphi | V} \braket{V | \theta} \braket{\theta | H} \braket{H | \varphi} \propto e^{-i(\chi + \varphi - \theta)}$. Note that the four bra-vectors in the scalar product are given by a cyclic permutation of the four ket-vectors. Hence, the particle exchange $\pi = (1\,2)(3\,4)$ in combination with the exchange of the roles of modes $2$ and $3$ in the two terms of the entangled state $\ket{\psi}$ (Eq.~\eqref{eq:inputState}, see also Fig.~\ref{fig:GenericScheme}, right panel) can be viewed as an effective four-cycle acted on the particles internal states (see also Eq.~\eqref{eq:overlaps}) which captures the dependence on the complex phases $\varphi, \chi,\theta$ (see also App.~\ref{app:kPcoherences} for a more detailed discussion).
Overall we find, for $k=4$,
\begin{equation}\label{eq:4pointcor_prediction}
\braket{ \hat{N}_1 \hat{N}_2 \hat{N}_3 \hat{N}_4}
= \frac{1}{8} \cos^2 \left(\frac{\phi}{2} \right),
\end{equation}
with collective 4P phase $\phi = \chi + \varphi - \theta$.
The full derivation is given in App.~\ref{app:kPcoherences}.
Thus, the 4P correlator oscillates with full interference contrast.
One can show that this observation also holds for arbitrary particle numbers $N>4$, without loss of contrast, by adding beam splitters in parallel and entangled pairs between the unentangled particles, as discussed in the beginning of this section. 

In contrast, for $k<N$, the structure of the input state in Eq.~\eqref{eq:inputState} and the specific interferometer topology ensure that all probed matrix elements $\braket{\bm{m} | \rho\inext^{(k)} | \bm{n} }$ in the expectation value~\eqref{eq:kpointcorrelator} are independent of the phases $\varphi,\, \chi,\, \theta$.
The idea is as follows: the three-point correlator only depends on the matrix elements of $\rho\inext^{(3)}$, see Eq.~\eqref{eq:kpointcorrelator}.
Note that the considered interferometer does not mix the modes $1,2$ with the modes $3,4$ and that the only non-vanishing $k$P coherences $\braket{\bm{m} | \rho\inext^{(k)} | \bm{n} }$ are indexed by $k$-tuples $\bm{m}$ and $\bm{n}$ that contain the same modes, only differing by a relative permutation $\bm{n} = \pi(\bm{m})$, as discussed after Eq.~\eqref{eq:kpointcorrelator}.
Therefore, the only contributing $3$P coherences in \eqref{eq:kpointcorrelator} for $k=3$ are --- up to complex conjugation and permuting both index tuples by the same permutation \footnote{Note that the reduced state $\rho\inext^{(k)}$ commutes with the action of the symmetric group, $\pi \rho\inext^{(k)} = \rho\inext^{(k)} \pi$. Therefore, its matrix elements are invariant under simultaneous permutations of both index-tuples $\bm{m}, \bm{n}$:  $\braket{\bm{m} | \rho\inext^{(k)} | \bm{n} } = \braket{\pi(\bm{m}) | \rho\inext^{(k)} | \pi(\bm{n})}$ for all  $\pi\in \mathrm{S}_k$ \cite{brunner_many-body_2019, brunner_many-body_2022}.} --- $\braket{123 |\, \tilde\rho \,| 213}$, $\braket{124 |\, \tilde\rho \,| 214}$, $\braket{134 |\, \tilde\rho \,| 143}$ and $\braket{234 |\, \tilde\rho \,| 243}$.
In all these terms only two particles are exchanged, similar to the $2$P HOM interference case (see Fig.~\ref{fig:GenericScheme}(a)), which only involves squared moduli of the internal states' overlaps. For example,
\begin{equation}
\braket{123 |\, \rho\inext^{(3)} \,| 213} \propto |\braket{\varphi | H} |^2 + |\braket{\varphi | V} |^2
\,.
\end{equation}
The first contribution comes from the first summation term of the initial state, where the particle in mode $2$ has horizontal polarisation, such that the exchange $\pi = (12)$ leads to $|\braket{\varphi | H} |^2$.
Correspondingly, the second contribution stems from the second term, where the particle in mode $2$ has vertical polarisation. Note that the cross contribution (which encodes the complex phase dependence for the four-point correlator as discussed before Eq.~\eqref{eq:4pointcor_prediction}) is zero since the particle in mode $3$ has orthogonal polarisation states in both terms.
Analogously, all other contributing matrix elements in Eq.~\eqref{eq:kpointcorrelator} are independent of the phases $\varphi, \theta, \chi$.
The same holds for $k=2$.
As a consequence, the $N$P collective phase remains invisible in all correlation orders lower than $N$.\\

Photonic experiments typically employ threshold-detectors and record coincidence events involving $k$ detectors, referred to as $k$-fold coincidences, where at least one photon is detected in each of the output ports $p_1,\dots, p_k$.
In contrast, measurements of $k$-point correlators, as in Eq.~\eqref{eq:kpointcorrelator}, are $k$P observables that require resolving the photon number at each detector, which is only possible with photon-number-resolving detectors \cite{Hadfield2009}.
In an $N$-particle scenario, the $N$-fold coincidence rate is equivalent to the $N$-point correlator, as necessarily only one photon impinges each detector. However, $k$-fold coincidence events with $k<N$ do not constitute true $k$P observables (as multiple photons may impinge on a single detector unnoticed) and are, in general, garnished by contributions from $l$P terms, with $k<l\leq N$, thus exhibiting weak traces of the $N$P collective phase $\phi$ (see App.~\ref{app:corr-coin}).

\section{Experiment}
\label{sec:experiment}

\begin{figure*}
  \includegraphics[width=\textwidth]{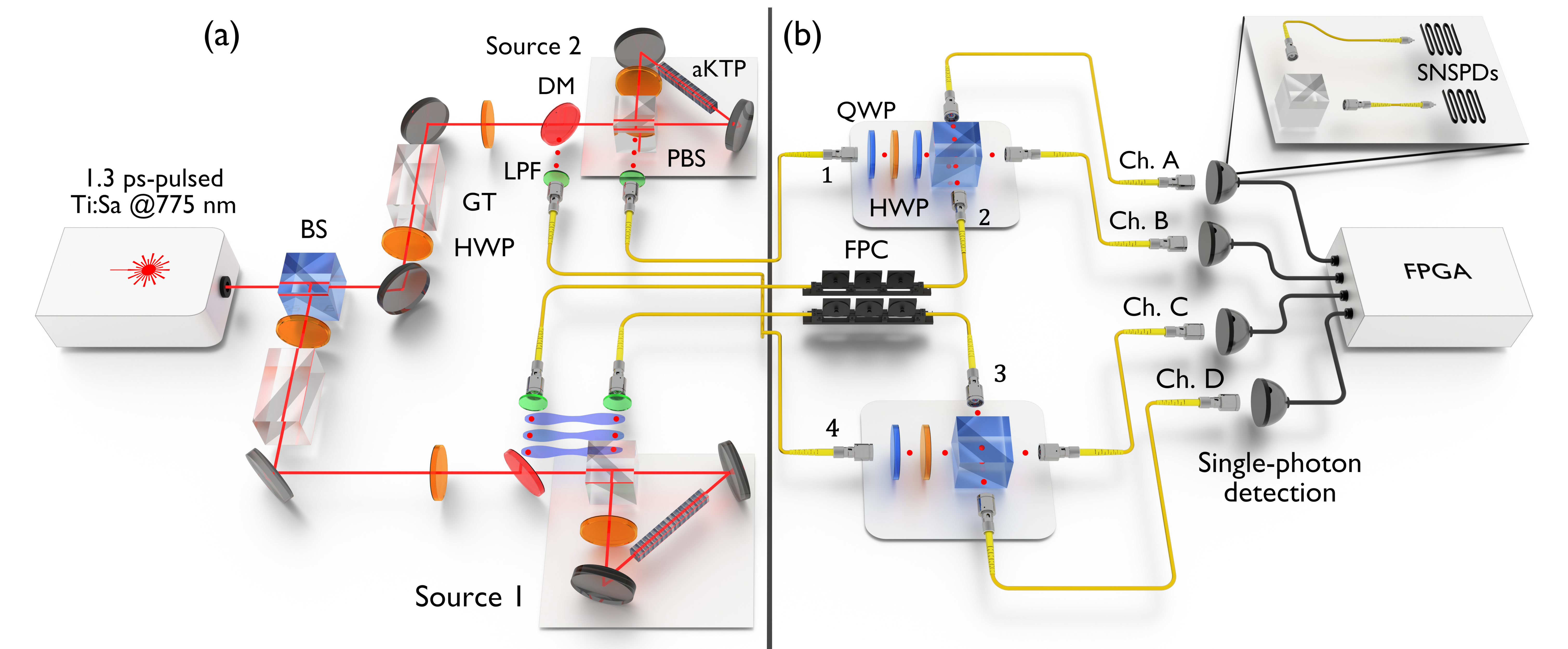}
  \caption{Experimental setup.
  \textbf{(a)} A $\qty{}{\pico\second}$-pulsed laser clocked at $\qty{80}{\mega\hertz}$ is split via a balanced beam splitter (BS) to pump two SPDC sources based on apodised KTP crystals (aKTP) embedded in Sagnac interferometers (Source 1 and Source 2).
  Half-wave plates (HWP) and Glan-Taylor polarisers (GT) are used to adjust the input power and set the pump polarisation to obtain entangled photon pairs (Source 1), or separable photon pairs (Source 2).
  Down-converted photons are separated from the laser light via polarising beam splitters (PBSs), dichroic mirrors (DMs), long-pass filters (LPFs), and collected through single-mode fibres.
  \textbf{(b)} Photons are adjusted in polarisation via combinations of HWPs and quarter-wave plates (QWP), or via fibre polarisation controllers (FPC), and guided to two independent balanced beam splitters.
  At the output, photons are detected by multiplexing each output channel (ch.A-ch.D) with two SNSPDs at the output ports of a PBS, as shown in the inset.
  The single-photon detection events are analysed at a field programmable gate array (FPGA) logic unit, and post-processed to determine every two-fold, three-fold, and four-fold coincidence event.}
  \label{fig:experimental setup}
\end{figure*}

\noindent \textbf{Experimental setup}\\
We experimentally tested the collective-phase interference as predicted by Eq.~\eqref{eq:4pointcor_prediction} by implementing the scheme in Fig.~\ref{fig:GenericScheme}(c) and the input state in Eq.~\eqref{eq:inputState}.
Figure \ref{fig:experimental setup} shows the experimental setup.
We produce the four-photon input states via two high-brightness, high-purity (spectral purity $\geq 98\%$), and high-fidelity telecom (emission wavelength \SI{1550}{nm}) spontaneous parametric down-conversion (SPDC) sources with apodised crystals \cite{graffitti2018independent,Pickston2021}, as schematically represented in Fig.~\ref{fig:experimental setup}(a).
In the employed Sagnac interferometer configuration \cite{Fedrizzi2007,Weston2016,Meraner2021}, we obtained experimental fidelities of $98.56(2)\%$ and $98.33(2)\%$ when adjusting the phase $\chi$ of the entangled photon pair in modes 2 and 3 of Eq.~\eqref{eq:inputState} to $\chi=0,\,\pi$ (representing the well-known Bell states $\ket{\psi^+}$ and $\ket{\psi^-}$), respectively, as shown in App.~\ref{appendix:state fidelity}.

Before interfering the photons as shown in Fig.~\ref{fig:experimental setup}(b), we prepare the polarisation states of the photons in modes 1 and 4 according to Eq.~\eqref{eq:separable state}, through combinations of quarter- and half-wave plates, and we set the phase $\chi$ of the entangled pair via fibre polarisation controllers.
More precisely, we varied the collective phase of the input state by changing the phase $\varphi$ of the photons in mode 1 and the phase $\chi$ of the entangled pair to obtain the Bell state $\ket{\psi^+}$ or $\ket{\psi^-}$, but fixing the phase $\theta=0$, corresponding to the setting of a diagonal polarisation state $\ket{D}_4=\hat{a}^\dagger_{4,S}(0)\ket{0}$ of the photons in mode 4.

After interfering, the photons collected at the output ports of the beam splitters are detected via superconducting nanowire single-photon detectors (SNSPDs) with quantum efficiencies of $\geq80\%$.
To account for the polarisation-dependent SNSPD detection efficiency, we project to the H/V-basis and multiplex the outputs to eight SNSPD channels.
The post-processing of the data retrieves the aggregated single counts of each output channel and all possible combinations of $k$-fold coincidences among these channels, with $k=2,\,3,\,4$.
The multiplexed eight-detector scheme has two additional advantages:
first, since SPDC sources are affected by multi-pair emission contributions (more than two photon pairs produced), this scheme makes it possible to reject coincidences of more than four photons, thus partially cleaning the four-fold statistics from these contributions.
Second, the presence of additional beam splitters and (non-number-resolving) detectors helps to provide a pseudo-photon-number resolution \cite{Heilmann2016}, which is required to perform correlation measurements as in Eq.~\eqref{eq:kpointcorrelator}.
Although a single multiplexing layer is insufficient for a complete correlation measurement, it already significantly reduces the visibility of the collective phase dependence of any $k(<4)$-fold coincidence events (see example in App.~\ref{app:corr-coin}).\\

\noindent \textbf{Experimental results}\\
The specific arrangement of SPDC sources in our experimental scheme causes double-pair emissions from an individual source to propagate through the setup with the same probability as the desired two-source emissions.
This results in a four-fold coincidence background from each source with the same order of magnitude as the coincidences from two-source emissions, thus reducing the visibility compared to the expected theoretical prediction of Eq.~\eqref{eq:4pointcor_prediction}.
However, our setup has the advantage of consisting of completely independent beam splitters, such that photons from single-source double-pair emissions do not interfere. The double-pair background 
is therefore independent of the collective phase by construction, and can be subtracted from the entanglement-induced four-photon collective interference phase signal.
Specifically, by blocking one of the two sources at a time, the four-photon background from the other source can be measured independent of the signal and subtracted from the raw data obtained in the main experiment \cite{Jones2020}.

We recorded photon events for combinations of the two phases $\chi=0,\,\pi$ and 31 settings of the phase $\varphi \in (-\pi/2,3\pi/2)$, thus testing the collective phase within the $3\pi$ phase range $\phi \in (-\pi/2, 5\pi/2)$.
We measured each setting of $\varphi$, at fixed $\chi$, over $\qty{60}{\s}$ and we averaged the measurements over 20 repetitions of this phase scan.
We subsequently performed the measurement of the background for each photon source by following the same procedure, but using a reduced set of 8 phase scans per source to obtain a good trade-off between acquiring sufficient data statistics and avoiding long-term drifts affecting the measurements.

\begin{figure}
    \centering
    \includegraphics[width=\linewidth]{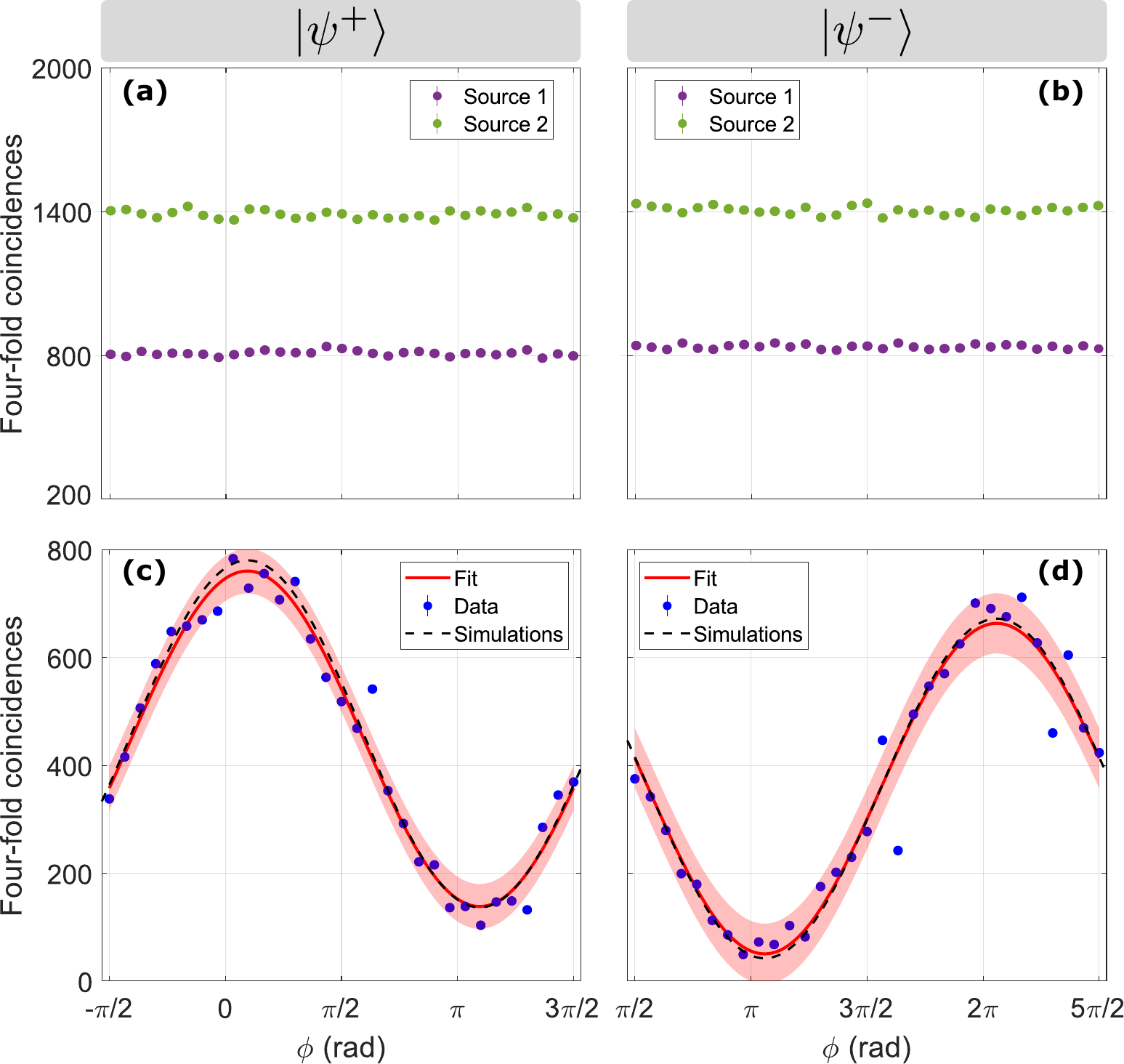}
    \caption{Results of four-fold coincidence counts for entanglement-induced collective $4$P interference. (a)(b) Four-fold background coincidences from multi-pair emissions of individual sources. (c)(d) Background-subtracted four-fold coincidences data (blue dots) fitted with a cosine function (red curve). The red-shaded region shows the fit prediction interval at a confidence level of one standard deviation. The graphs include the results of the multi-photon interference simulations (black dashed curve). The visibility of the fit is $69.2(2.5)\%$ and $85.8(4.5)\%$  for panels (c) and (d), respectively. Simulations (App.~\ref{appendix:simulations}) predict a visibility of $70.2(1.2)\%$ and $88.1(1.7)\%$. The integration time for each point of all panels is \qty{60}{\second}.}
    \label{fig:background subtraction}
\end{figure}

Figure \ref{fig:background subtraction} shows the results of both the four-fold background coincidences and the background-subtracted four-fold coincidences when preparing the photon pair in modes 2 and 3 in the two Bell states $\ket{\psi^+}$ ($\chi=0$) and $\ket{\psi^-}$ ($\chi=\pi$).
The background coincidences in Figs.~\ref{fig:background subtraction}(a) and \ref{fig:background subtraction}(b) have negligible fluctuations with relative magnitude $\simeq 1.2\%$, uncorrelated to the collective phase $\phi$. 
In contrast, the background-subtracted four-fold coincidences in Figs.~\ref{fig:background subtraction}(c) and \ref{fig:background subtraction}(d) exhibit pronounced cosine oscillations, as predicted in Eq.~\eqref{eq:4pointcor_prediction}, with high visibilities of $69.2(2.5)\%$ and $85.8(4.5)\%$, respectively.
This distinctive dependence of four-fold coincidences on the collective phase cannot be attributed to fluctuations of single counts or two-fold and three-fold coincidences, which exhibit only weak fluctuations imputable to secondary effects, as detailed in App.~\ref{appendix:raw data}, where we also present the raw (uncorrected) four-photon coincidence data.
Moreover, the experimental data are in good agreement with the numerical simulations of the experiment, which provide visibilities of $70.2(1.2)\%$ and $88.1(1.7)\%$, respectively.
The numerical results are obtained by taking into account input and output losses, unbalanced splitting ratios of the beam splitters, multi-pair emissions of the sources up to six photons, and laser power drifts (see App.~\ref{appendix:simulations}).

These numerical simulations also help to quantify the contribution of effects causing the residual visibility reduction of four-fold coincidences in Fig.~\ref{fig:background subtraction}(d) and, in particular, Fig.~\ref{fig:background subtraction}(c).
The primary factors affecting the visibility are, in order of significance: laser power drifts, higher-order ($>3$) photon-pair emissions, and the imperfect implementation of the unitary transformation.
Deviations in the average laser power during the main measurement and the two background measurements in Figs.~\ref{fig:background subtraction}(a) and \ref{fig:background subtraction}(b) lead to the subtraction of a background level that differs from the level affecting the main measurement.
The background four-fold coincidences in Fig.~\ref{fig:background subtraction}(a) are slightly lower than those measured in Fig.~\ref{fig:background subtraction}(b). This causes the slightly lower visibility in Fig.~\ref{fig:background subtraction}(c) with respect to the measurement in Fig.~\ref{fig:background subtraction}(d), which is virtually unaffected by power drifts (see simulation in App.~\ref{appendix:power drift} for details).
The second noise contribution is associated with higher-order emissions with at least three photon pairs emitted by the two sources---two pairs in one source, and one in the other. Our background subtraction procedure cannot account for this process.
Imperfections in the experimental implementation of the unitary transformation are mainly caused by the slight manufacturing deviations of beam splitters from a 50:50 splitting ratio.
Additional minor contributions include residual polarisation dependence of the beam splitters' splitting ratios, the imperfect spectral purity of the down-converted photons or slight temporal indistinguishability at the beam splitters, as well as imperfect state preparation.

\section*{Conclusions}

We have shown in theory and photonic experiments that the entanglement between particles at the input ports of independent and separate beam splitters sets the stage for a collective interference effect of all particles, which cannot be traced back to the interference 
of smaller subsets.
The peculiarity of this phenomenon lies in the fact that a collective many-body dynamics arises despite the setup allowing only for pairwise interference and some particles occupying distinguishable internal states.
The entanglement in the input photon state bridges the gaps between the disjoint interferometers and, therefore, acts as a mediator of the collective interference in a nonlocal fashion.
Compared to the conventional case of non-entangled photons \cite{Shchesnovich2018,Jones2020}, this effect has two fundamental advantages for scaling towards more particles: first, the equivalent interference scenario with $N>4$ photons can be implemented by adding beam splitters and pairs of entangled photons in parallel, without the requirement of engineering additional internal dof as $N$ increases, such that the single-particle Hilbert space dimension stays constant at two \cite{brunner_many-body_2019}. Second, the visibility of the entanglement-induced collective phase interference is ideally always one and does not diminish with a growing number of particles.
This will be instrumental in obtaining clear signals in interference scenarios with a large number of particles involved.

Our experimental data and numerical simulations produce a high interference visibility, which is mainly limited by the nature of the SPDC photon sources and the errors introduced with the background subtraction.
However, for $N>4$ input particles, the partial (input-output) connectivity of the interferometer necessitates the participation of all $N/2$ sources to achieve an $N$-fold coincidence, that is, background subtraction is no longer required to obtain visibilities comparable with the highest visibility obtained here ($\sim 86\%$).
The detrimental influence of multi-pair emissions could also be completely overcome by using deterministic single-photon sources, such as semiconductor quantum dots, if the required input state can be engineered by appropriate excitation schemes \cite{HuberStrainTunableQD2018,SbresnySTIMTPE2022,karli2023controlling} and high photon indistinguishability between separate source can be achieved \cite{Zhai2022}.  

Our study gives another striking example of how entanglement can play a key role in shaping interference phenomena \cite{Pittman1996,Kim2005,Agne2016}. It reveals a hitherto unexplored interference effect that, by virtue of the partial distinguishability of the involved particles, extends the complexity of traditional many-body interference scenarios, potentially leading to unexpected phenomena.
For instance, a recent work has shown that by using $N \geq 7$ partially distinguishable particles, indistinguishable bosons do not maximise the probability of bunching in a subset of output modes \cite{Seron2023}.
This prompts questions on whether and how a collective behaviour, as induced here through entanglement between internal dof of a subset of particles, can affect the bunching dynamics in similar many-body conditions.
The interference scheme presented here readily extends to larger entangled GHZ states while preserving its properties. Employing such multi-partite entangled states offers various advantages in quantum communication protocols \cite{Hillery1999,Murta2020}. Consequently, it seems worthy to further investigate the potential of entanglement-induced collective interference with GHZ states for multiparty quantum communication.
Furthermore, the presence of an $N$P collective phase producing high interference contrast could be exploited for quantum metrology purposes. In particular, the possibility of scaling up the scheme by adding only Bell pairs can be beneficial to avoid the typical complex preparation of NOON states and their exposure to decoherence at large photon numbers $N$ \cite{Thekkadath2020}.
Overall, the combination of partial distinguishability and entanglement in many-body interference could serve as a valuable tool for advancing quantum technologies, or manifest as a subtle yet noteworthy side-effect.
\vspace{0.5cm}

\section*{Acknowledgements}

\noindent The authors acknowledge Stefan Frick for valuable discussions during the preparation of the manuscript. T.F., G.W. and R.K. acknowledge the Austrian Science Fund (FWF) projects FG 5, F 7114 and W1259 (DK-ALM) for financial support. C.D. acknowledges the Georg H. Endress Foundation for support and the Freiburg Institute for Advanced Studies for a FRIAS Junior Fellowship. A.F. is supported by the EPSRC Quantum Technology Hub in Quantum Communication (EP/T001011/1).

\bibliography{references.bib}

\appendix

\section{Four-point correlator and collective phase}
\label{app:kPcoherences}

In the following, we derive Eq.~\eqref{eq:4pointcor_prediction}.
The expectation value of the four-point correlator can be evaluated as $\braket{ \hat{N}_1 \hat{N}_2 \hat{N}_3 \hat{N}_4} = \braket{\psi | \mathcal{U}^\dagger \hat{N}_1 \hat{N}_2 \hat{N}_3 \hat{N}_4 \mathcal{U} | \psi}$, with $\mathcal{U}=U^{\otimes N}$ and $U$ the single-particle unitary transformation implemented by the array of beam splitters in Fig.~\ref{fig:GenericScheme}(c):
\begin{equation}
    \label{eq:unitary}
    U=\frac{1}{\sqrt{2}}
    \begin{pmatrix}
    1 & 1 & 0 & 0\\
    1 & -1 & 0 & 0\\
    0 & 0 & 1 & 1\\
    0 & 0 & 1 & -1
    \end{pmatrix}.
\end{equation}
For convenience, we introduce 2P operators $\hat{A}_{ij, kl} = \sum_{\alpha, \beta} \hat{a}^\dagger_{i\alpha} \hat{a}^\dagger_{j\beta} \hat{a}_{k\alpha} \hat{a}_{l\beta}$, where the sum runs over a complete set of basis modes of the internal Hilbert space, e.g. $\alpha, \, \beta \in \lbrace H, V \rbrace$, in our situation.
Since the unitary in Eq.~\eqref{eq:unitary} does not mix the modes $(1,2)$ with the modes $(3,4)$, we can look at the measurements on both subsets of modes individually.
We can write $\mathcal{U}^\dagger \hat{N}_1 \hat{N}_2 \hat{N}_3 \hat{N}_4 \mathcal{U} = \mathcal{U}^\dagger \hat{N}_1 \hat{N}_2 \mathcal{U} \mathcal{U}^\dagger \hat{N}_3 \hat{N}_4 \mathcal{U}$ and calculate
\begin{equation}
\begin{split}
& \mathcal{U}^\dagger \hat{N}_1 \hat{N}_2 \mathcal{U} \\
&= \frac{1}{4} \left( \hat{A}_{12,12} - \hat{A}_{12,21} - \hat{A}_{21,12} + \hat{A}_{21,21} + \hat{R}_{12} \right)
\end{split}
\end{equation}
where $\hat{R}_{12}$ collects those terms $\hat{A}_{ij,kl}$ with either $i=j = 1,2$ or $k=l = 1,2$.
These terms have zero overlap with the initial state $\ket{\psi}$ given in Eq.~\eqref{eq:inputState}.
Analogously, we have
\begin{equation}
\begin{split}
& \mathcal{U}^\dagger \hat{N}_3 \hat{N}_4 \mathcal{U} \\
&= \frac{1}{4} \left(\hat{A}_{34,34} - \hat{A}_{34,43} - \hat{A}_{43,34} + \hat{A}_{43,43} + \hat{R}_{34} \right) \,,
\end{split}
\end{equation}
where, as before, $\hat{R}_{34}$ describes the terms that have no overlap with $\ket{\psi}$.
Because $\hat{A}_{ij,kl}$ contains a sum over internal indices, it is symmetric under permuting the first and second tuples of indices $\hat{A}_{ij,kl} = \hat{A}_{ji, lk}$. By omitting terms involving $\hat{R}_{12}, \, \hat{R}_{34}$, we have
\begin{equation}\label{eq:deriv_app_fourpointcor}
\begin{split}
& \mathcal{U}^\dagger \hat{N}_1 \hat{N}_2 \hat{N}_3 \hat{N}_4 \mathcal{U} \\
&= \frac{1}{16} \left(2 \hat{A}_{12,12} - 2 \hat{A}_{12,21}) (2 \hat{A}_{34,34} - 2 \hat{A}_{34,43}) + \dots \right) \\
&= \frac{1}{4} \Big( \hat{A}_{1234,1234} - \hat{A}_{1234,1243} \\
&\qquad\qquad - \hat{A}_{1234,2134} + \hat{A}_{1234, 2143} + \dots \Big) \,.
\end{split}
\end{equation}
The 4P operators $\hat{A}_{ijkl,mnop} = \sum_{\alpha, \beta, \gamma, \delta} \hat{a}^\dagger_{i\alpha} \hat{a}^\dagger_{j\beta} \hat{a}^\dagger_{k\gamma} \hat{a}^\dagger_{l\delta} \hat{a}_{m\alpha} \hat{a}_{n\beta} \hat{a}_{o\gamma} \hat{a}_{p\delta}$ are defined analogously to their 2P counterparts.
The first operator results in a contribution where no particles are exchanged and, hence, is independent of the internal states of the particles.
The second and third terms describe the exchange of particles in modes $1,2$ and $3,4$, respectively. As we will see below, these contributions depend only on the squared moduli of the internal states' overlaps and are, therefore, independent of any phase.
Only the fourth operator $\hat{A}_{1234, 2143}$ (realising a simultaneous exchange of particles in $1,2$ and in $3,4$) mediates the collective phase dependence.

As a next step, we need to evaluate all four non-zero contributions in the initial state \eqref{eq:inputState}, which is a sum of two states
\begin{equation}\label{eq:initial_state_appendix}
\begin{split}
&\ket{\psi} = \frac{1}{\sqrt{2}} (\ket{\psi_1} + \ket{\psi_2} ) \\
& \ket{\psi_1} = \hat{a}^\dagger_{1S}(\varphi) \hat{a}^\dagger_{2H} \hat{a}^\dagger_{3V} \hat{a}^\dagger_{4 S}(\theta) \ket{0} \\
& \ket{\psi_2} = e^{-i \chi} \hat{a}^\dagger_{1S}(\varphi) \hat{a}^\dagger_{2V} \hat{a}^\dagger_{3H} \hat{a}^\dagger_{4 S}(\theta) \ket{0} \,.
\end{split}
\end{equation}
We start with $A_{1234,2143}$:
\begin{equation}\label{eq:app1_helper1}
\begin{split}
&\braket{\psi_1 | \hat{A}_{1234, 2143} | \psi_1} = \braket{\varphi | H} \braket{H | \varphi} \braket{V | \theta} \braket{\theta | V} = 1/4 \\
&\braket{\psi_2 | \hat{A}_{1234, 2143} | \psi_2} = \braket{\varphi | V} \braket{V | \varphi} \braket{H | \theta} \braket{\theta | H} = 1/4 \\
&\braket{\psi_1 | \hat{A}_{1234, 2143} | \psi_2} = e^{-i\chi} \braket{\varphi | V} \braket{H | \varphi} \braket{V | \theta} \braket{\theta | H} \\
&= e^{-i\chi} \braket{\varphi | V} \braket{V | \theta} \braket{\theta | H} \braket{H | \varphi} = \frac{1}{4} e^{-i\chi -i\varphi + i\theta} \\
&\braket{\psi_2 | \hat{A}_{1234, 2143} | \psi_1} = \braket{\psi_1 | \hat{A}_{1234, 2143} | \psi_2}^\ast \,.
\end{split}
\end{equation}
The fourth line is a reordering of the terms of the third line that shows that the four bra-vectors in the scalar products are a cyclic permutation of the four ket-vectors. One can interpret this effective four-cycle permutation stemming from the combination of exchanging particles in modes $(1,2)$ and $(3,4)$ (the effect of the operator $\hat{A}_{1234,2143}$) and exchanging the roles of modes $2$ and $3$ ($\ket{\psi_1}$ versus $\ket{\psi_2}$, see Eq.~\eqref{eq:initial_state_appendix}).
The last line follows from $\braket{\psi_2 | \hat{A}_{1234, 2143} | \psi_1} = \braket{\psi_1 | \hat{A}_{3412, 4321} | \psi_2}^\ast = \braket{\psi_1 | \hat{A}_{1234, 2143} | \psi_2}^\ast$, where the second equality follows by permuting the first and second index tuples by the permutation $\pi = (1\,3)(2\,4)$.
Altogether we end up with
\begin{equation}
\begin{split}\label{eq:deriv_app_fourpointcor1}
&\braket{\psi | \hat{A}_{1234, 2143} | \psi} = \frac{1}{4} \left( 1 + \cos( \theta - \chi - \varphi) \right)\\
&= \frac{1}{4} \left( 1 + \cos( \chi + \varphi - \theta) \right) \,.   
\end{split}
\end{equation}
Next we calculate $\hat{A}_{1234, 1243}$: here only the particles in modes $3$ and $4$ are exchanged. The expectation values in the states $\ket{\psi_1}$ and $\ket{\psi_2}$ are readily calculated to be
\begin{equation}
\begin{split}\label{eq:app1_helper2}
&\braket{\psi_1 | \hat{A}_{1234,1243} | \psi_1} 
= \braket{\varphi | \varphi} \braket{H | H} \braket{V | \theta} \braket{\theta | V} = \frac{1}{2} \\
&\braket{\psi_2 | \hat{A}_{1234,1243} | \psi_2} 
= \braket{\varphi | \varphi} \braket{V | V} \braket{H | \theta} \braket{\theta | H} = \frac{1}{2} \,.
\end{split}
\end{equation}
Permuting only particles in mode $3,4$ leads to absolute squares of $\braket{V | \theta}$ and $\braket{H | \theta}$, which do not depend on the complex phase. Interestingly, the cross-terms vanish, since the permutation of the particles in mode $3,4$ enforces an inner product of the orthogonal states $V$ and $H$:
\begin{equation}
\begin{split}
&\braket{\psi_1 | \hat{A}_{1234, 1243} | \psi_2} = \frac{1}{2} e^{-i\chi} \braket{\varphi | \varphi} \braket{H | V} \braket{V | \theta} \braket{\theta | H} = 0 \\
& \braket{\psi_2 | \hat{A}_{1234, 1243} | \psi_1} = \braket{\psi_1 | \hat{A}_{1234, 1243} | \psi_2}^\ast = 0 \,.
\end{split}
\end{equation}
An analogous calculation for the exchange of particles in modes 1 and 2 through $\hat{A}_{1234, 2134}$ leads to similar results.
With this we end up with
\begin{equation}\label{eq:deriv_app_fourpointcor2}
\braket{\psi | \hat{A}_{1234, 1243} | \psi} = \braket{\psi | \hat{A}_{1234, 2134} | \psi} = \frac{1}{2} \left(\frac{1}{2} + \frac{1}{2} \right) = \frac{1}{2} \,.
\end{equation}
For the last contribution, $\hat{A}_{1234, 1234}$, we reorder $\hat{A}_{1234,1234} = \hat{N}_1 \hat{N}_2 \hat{N}_3 \hat{N}_4$ and find
\begin{equation}\label{eq:deriv_app_fourpointcor3}
\braket{\psi | \hat{A}_{1234, 1234} | \psi} = 1 \,,
\end{equation}
since each mode is occupied by exactly one particle.

We combine Eqs.~\eqref{eq:deriv_app_fourpointcor}, \eqref{eq:deriv_app_fourpointcor1}, \eqref{eq:deriv_app_fourpointcor2} and \eqref{eq:deriv_app_fourpointcor3} and obtain the expectation value of the ideal four-point correlator:
\begin{equation}
\begin{split}
&\braket{ \hat{N}_1 \hat{N}_2 \hat{N}_3 \hat{N}_4} = \braket{\psi | \mathcal{U}^\dagger \hat{N}_1 \hat{N}_2 \hat{N}_3 \hat{N}_4 \mathcal{U} | \psi} \\
&= \frac{1}{4} \left( 1 - \frac{1}{2} - \frac{1}{2} + \frac{1}{4}\left( 1 + \cos(\chi + \varphi - \theta) \right) \right) \\
&= \frac{1}{16} \left( 1 + \cos(\chi + \varphi - \theta) \right) = \frac{1}{8} \cos^2 \left( \frac{\chi + \varphi - \theta}{2} \right),
\end{split}
\end{equation}
as given in Eq.~\eqref{eq:4pointcor_prediction}.

\section{Correlation versus coincidence measurements} \label{app:corr-coin}

The operator associated with the measurement of \emph{at least one} photon in output mode $p$ is given by (disregarding internal degrees of freedom)
\begin{align}\label{eq:1pop}
    \hat{M}_p&=\sum_{n=1}^\infty \frac{(-1)^{n+1}}{n!} (\hat{a}_p^\dagger)^n \hat{a}_p^n\\
    &=\sum_{n=1}^\infty \frac{(-1)^{n+1}}{n!} \hat{N}_p(\hat{N}_p-1)\dots (\hat{N}_p-n+1).
\end{align}
Indeed, the number state $\ket{m}$ with $m$ photons in mode $p$ satisfies
\begin{align}\label{eq:1pop2}
    (\mathbb{I}-\hat{M}_p)\ket{m}&=\sum_{n=0}^\infty \frac{(-1)^{n}}{n!} (\hat{a}_p^\dagger)^n\hat{a}_p^n \ket{m}\\
    &=\sum_{n=0}^m \frac{m!}{n!(m-n)!} (-1)^{n} \ket{m}\\
    &=\delta_{m,0}\ket{m}, 
\end{align}
where we have made use of the binomial formula.
Therefore, $k$-fold coincidence rates are obtained as the expectation value of 
\begin{align}\label{eq:coinc}
    \hat{M}_{p_1}\dots \hat{M}_{p_k}.
\end{align}
The lowest-order term in the expansion of Eq.~\eqref{eq:coinc} is the $k$-point correlator $\hat{N}_{p_1} \dots \hat{N}_{p_k}$, corresponding to taking terms with $n=1$ in Eq.~\eqref{eq:1pop}.
This leading term is corrected by higher-order observables, which are sensitive to interference from higher orders than $k$.

As an example, when considering a system of $N=4$ particles and neglecting terms beyond the fourth order, the explicit expansion of Supplementary Eq.~\eqref{eq:coinc} reads
\begin{widetext}
    \begin{align}
    \hat{M}_{p_1} &= \hat{N}_{p_1} -\frac{1}{2}\hat{N}_{p_1}(\hat{N}_{p_1}-1) +\frac{1}{6}\hat{N}_{p_1}(\hat{N}_{p_1}-1) (\hat{N}_{p_1}-2)- \frac{1}{24}\bm{\hat{N}_{p_1}(\hat{N}_{p_1}-1) (\hat{N}_{p_1}-2) (\hat{N}_{p_1}-3)} \label{eq:Mp1}\\
     \hat{M}_{p_1}\hat{M}_{p_2} &=  \hat{N}_{p_1} \hat{N}_{p_2}-\frac{1}{2}\hat{N}_{p_1}(\hat{N}_{p_1}-1)\hat{N}_{p_2}-\frac{1}{2}\hat{N}_{p_1}\hat{N}_{p_2}(\hat{N}_{p_2}-1) +\frac{1}{4}\bm{\hat{N}_{p_1}(\hat{N}_{p_1}-1)\hat{N}_{p_2}(\hat{N}_{p_2}-1)}\notag \\&\qquad +\frac{1}{6}\bm{\hat{N}_{p_1}(\hat{N}_{p_1}-1)(\hat{N}_{p_1}-2)\hat{N}_{p_2}}+\frac{1}{6}\bm{\hat{N}_{p_1}\hat{N}_{p_2}(\hat{N}_{p_2}-1)(\hat{N}_{p_2}-2)} \label{eq:Mp1Mp2}\\
\hat{M}_{p_1}\hat{M}_{p_2}\hat{M}_{p_3} &=  \hat{N}_{p_1} \hat{N}_{p_2} \hat{N}_{p_3}-\frac{1}{2}\bm{\hat{N}_{p_1}(\hat{N}_{p_1}-1)\hat{N}_{p_2}\hat{N}_{p_3}}-\frac{1}{2}\bm{\hat{N}_{p_1}\hat{N}_{p_2}(\hat{N}_{p_2}-1)\hat{N}_{p_3}}-\frac{1}{2}\bm{\hat{N}_{p_1}\hat{N}_{p_2}\hat{N}_{p_3}(\hat{N}_{p_3}-1)} \label{eq:Mp1Mp2Mp3}\\
\hat{M}_{p_1}\hat{M}_{p_2}\hat{M}_{p_3}\hat{M}_{p_4} &=  \bm{\hat{N}_{p_1} \hat{N}_{p_2} \hat{N}_{p_3}\hat{N}_{p_4}}, \label{eq:Mp1Mp2Mp3Mp4}
\end{align}
\end{widetext}
where fourth-order terms are highlighted.
\begin{figure}
\centering
\includegraphics[width=\linewidth]{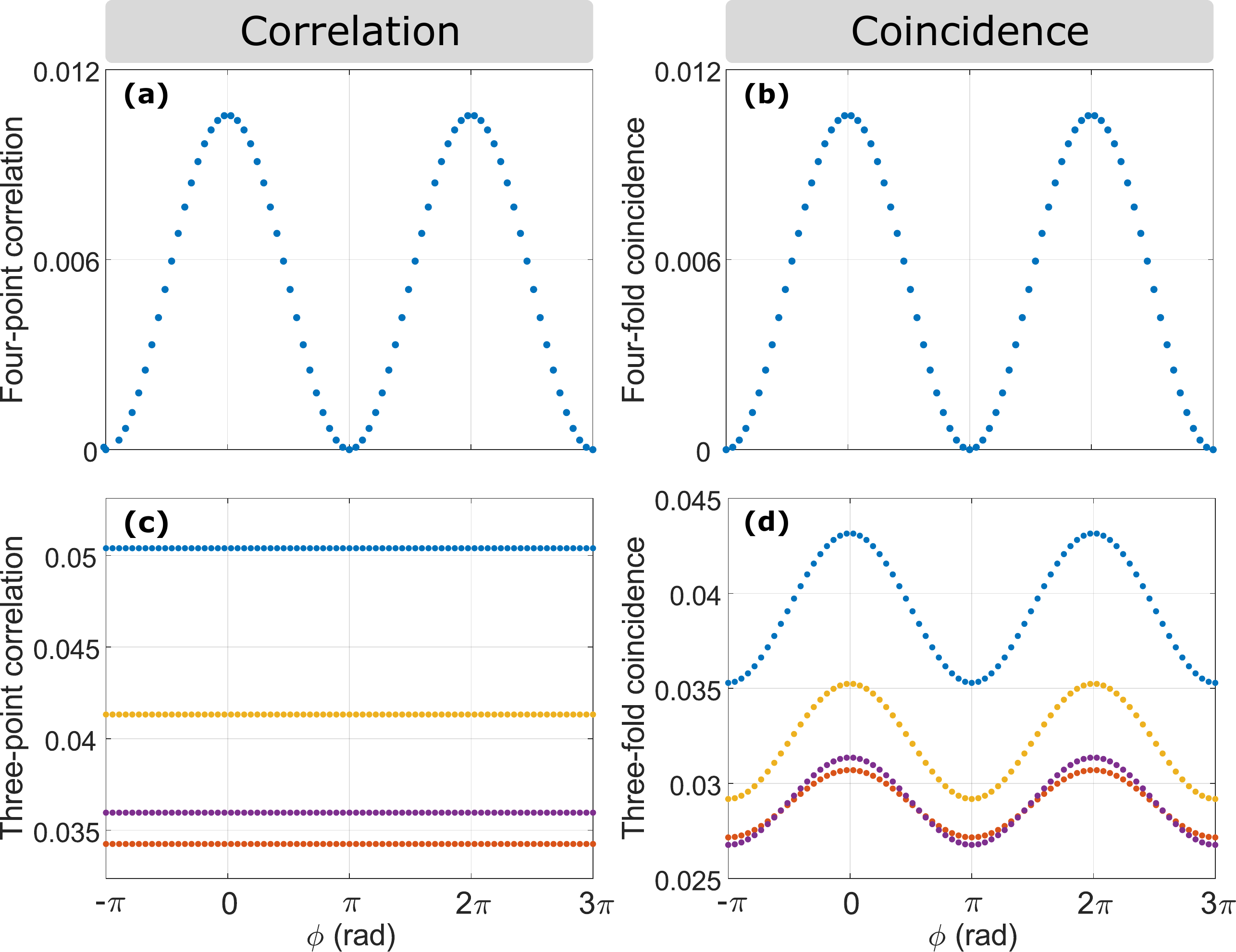}
\caption{Simulations of correlations and coincidence events. Correlations show \textbf{(a)} a fully visible fringe pattern of the four-point correlator, whereas \textbf{(c)} the three-point correlators, which depend on the coherence of the reduced three-particle states, are all independent of the collective phase.
We sampled input and output transmission efficiencies from a normal distribution with a mean value of 0.8 and a standard deviation of 0.1.
Different total losses displace the curves among each other, with higher losses resulting in lower curves.
Simulations of coincidence events show \textbf{(b)} four-fold coincidence with same behaviour as the four-point correlation (see Eq.~\eqref{eq:Mp1Mp2Mp3Mp4}), and \textbf{(d)} a signature of the collective 4P phase also in the three-fold coincidences, as these are not proper 3P measurements.
The visibility of this signature strongly depends on the optical losses, with lower values as losses increase.}
\label{fig:coincidences vs correlations}
\end{figure}
The presence of fourth-order terms in Eq.~\eqref{eq:Mp1}-\eqref{eq:Mp1Mp2Mp3} results in a residual dependence of $k(<N)$-fold coincidences on the collective phase $\phi$, where the magnitude of this influence increases as $k$ approaches $N$.
The presence of these higher-order contributions is illustrated in Fig.~\ref{fig:coincidences vs correlations}: numerical simulations indeed show that three-fold coincidences depend on $\phi$, but with much weaker visibility than for the four-fold coincidence.

To grasp the dependence of three-fold coincidences on the collective phase $\phi$, consider the three-fold events among, for example, the output channels A, C, and D in the lossless conditions where $\phi=\pi, \, 0$. The output mode-occupation lists (counting the number of particles in each output mode) contributing to such events are $\mathbf{s_1}=(2, \, 0, \, 1, \, 1)$ and $\mathbf{s_2}=(1, \, 1, \, 1, \, 1)$. When $\phi=\pi$, the absence of four-fold coincidences indicates a bunching behaviour of photons, resulting in the suppression of events corresponding to $\mathbf{s_2}$ and in a three-fold coincidence probability solely given by $\mathbf{s_1}$, i.e., $p(\mathrm{ACD}|\phi=\pi)=p(\mathbf{s_1}|\phi=\pi)$. For symmetry reasons, $\mathbf{s_1}$ must have the same probability as the output state $\mathbf{s_3}=(0, \, 2, \, 1, \, 1)$ (which does not produce a coincidence in A, C and D). 
When $\phi=0$, all three states $\mathbf{s_1}$, $\mathbf{s_2}$, and $\mathbf{s_3}$ are allowed, of which $\mathbf{s_2}$ can be achieved via two distinct exchange processes at the beam splitter connecting modes A and B (double reflection and double transmission), while $\mathbf{s_1}$ and $\mathbf{s_3}$ can only be obtained via a single process each. Intuitively, this halves the probability of $\mathbf{s_1}$ with respect to the $\phi=\pi$-scenario, whereas the probability of $\mathbf{s_2}$ is double the probability of $\mathbf{s_1}$, that is, $p(\mathbf{s_2}|\phi=0)=p(\mathbf{s_1}|\phi=\pi)$. Therefore, the three-fold coincidence probability becomes $p(\mathrm{ACD}|\phi=0)=p(\mathbf{s_1}|\phi=0)+p(\mathbf{s_2}|\phi=0)=p(\mathbf{s_1}|\phi=\pi)(\frac{1}{2}+1)=\frac{3}{2}p(\phi=\pi)$. When losses are considered, such as in Fig. 4, the factor $3/2$ is reduced because three-fold coincidences from $\mathbf{s_1}$ are less affected by losses with respect to $\mathbf{s_2}$ (one photon loss in channel A of $\mathbf{s_1}$ still result in a coincidence).

Simulations of coincidences with the addition of one detector multiplexing layer, as in Fig.~\ref{fig:experimental setup}, predict a further visibility reduction of a factor $\sim 2.3$ (influenced by the value of the total optical losses) in the three-fold coincidences fluctuations of Fig.~\ref{fig:coincidences vs correlations}(d).\\

\begin{figure*}
    \centering
    \includegraphics[width=0.8\textwidth]{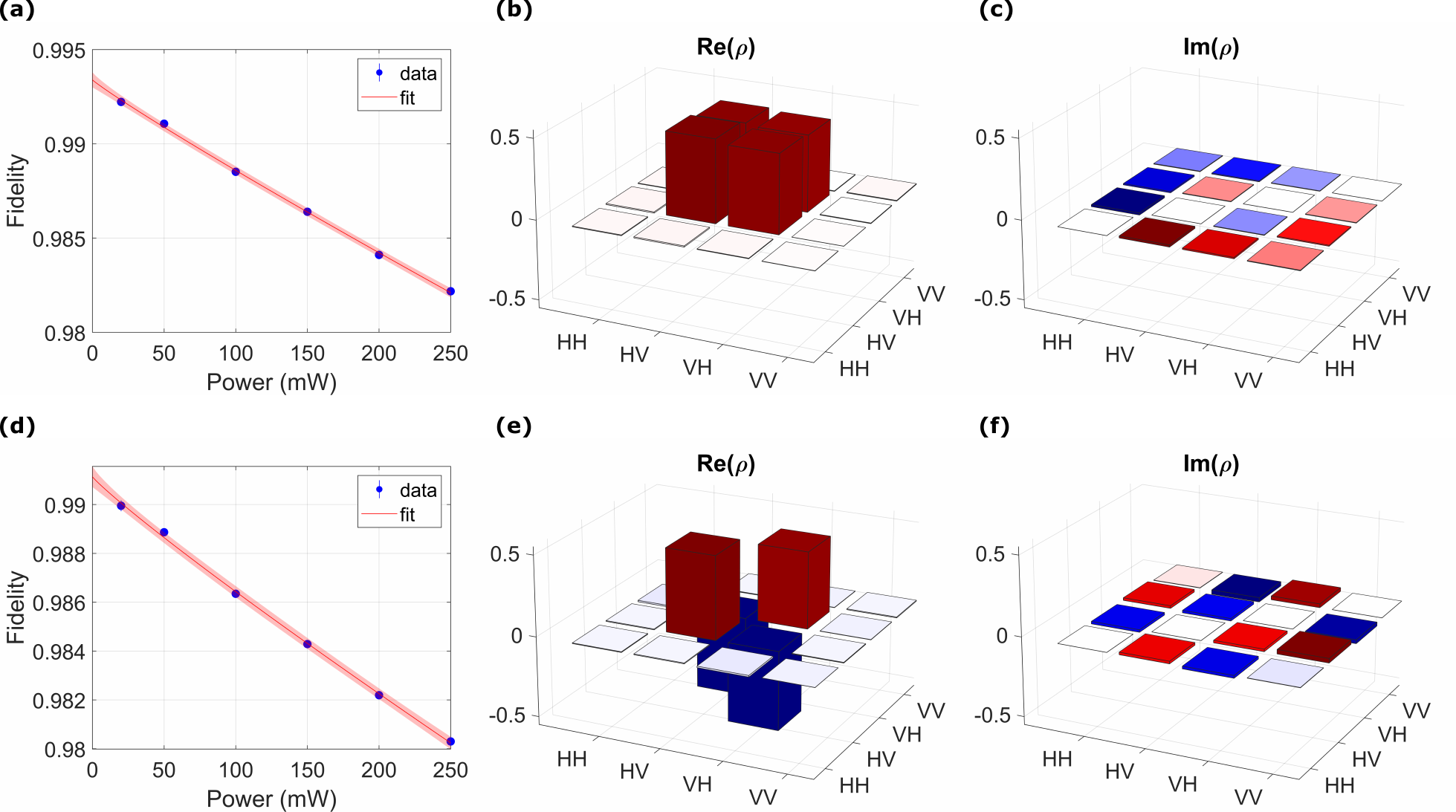}
    \caption{Tomography of the entangled photon pair. The panels in the upper and lower rows are the results for the input states $\ket{\psi^+}$ and $\ket{\psi^-}$ of the photons in modes 2 and 3, respectively. Panels \textbf{(a)} and \textbf{(d)} show the dependence of the states' fidelities on the laser pump power, whereas the other panels show the real (\textbf{(b)} and \textbf{(e})) and imaginary (\textbf{(c)} and \textbf{(f)}) parts of the reconstructed density matrix $\rho$ for a representative value of \qty{150}{\milli\watt}.}
    \label{fig:tomography}
\end{figure*}

\section{State fidelity} \label{appendix:state fidelity}

We evaluated the state overlap, referred to as fidelity \cite{Jozsa1994}, of the entangled state of the photons in modes 2 and 3 with respect to the Bell states $\ket{\psi^+},\,\ket{\psi^-}$, corresponding to setting $\chi=0,\,\pi$ in Eq.~\eqref{eq:inputState}, by performing a maximum likelihood state tomography \cite{Altepeter2005}.
To this end, the source outputs are connected to two polarisation-tomography stages where we measure single counts and coincidence events.
As mentioned in the main text, the phase $\chi$ is adjusted via fibre polarisation controllers.
We performed six state tomographies for the average pump power range \qty{20}{\milli\watt}--\qty{250}{\milli\watt}, which includes the power used to perform the experiment $\simeq \qty{174}{\milli\watt}$, and we evaluated the fidelity of the resulting density matrices with the corresponding states $\ket{\psi^+},\,\ket{\psi^-}$.

The results of these measurements are shown in Fig.~\ref{fig:tomography}.
From the fit of the data, we find a fidelity at $\qty{174}{\milli\watt}$ of 98.53(2)\% and 98.33(2)\%, respectively, for $\ket{\psi^+}$ and $\ket{\psi^-}$.

\section{Raw data} \label{appendix:raw data}

\begin{figure*}
    \centering
    \includegraphics[width=0.65\textwidth]{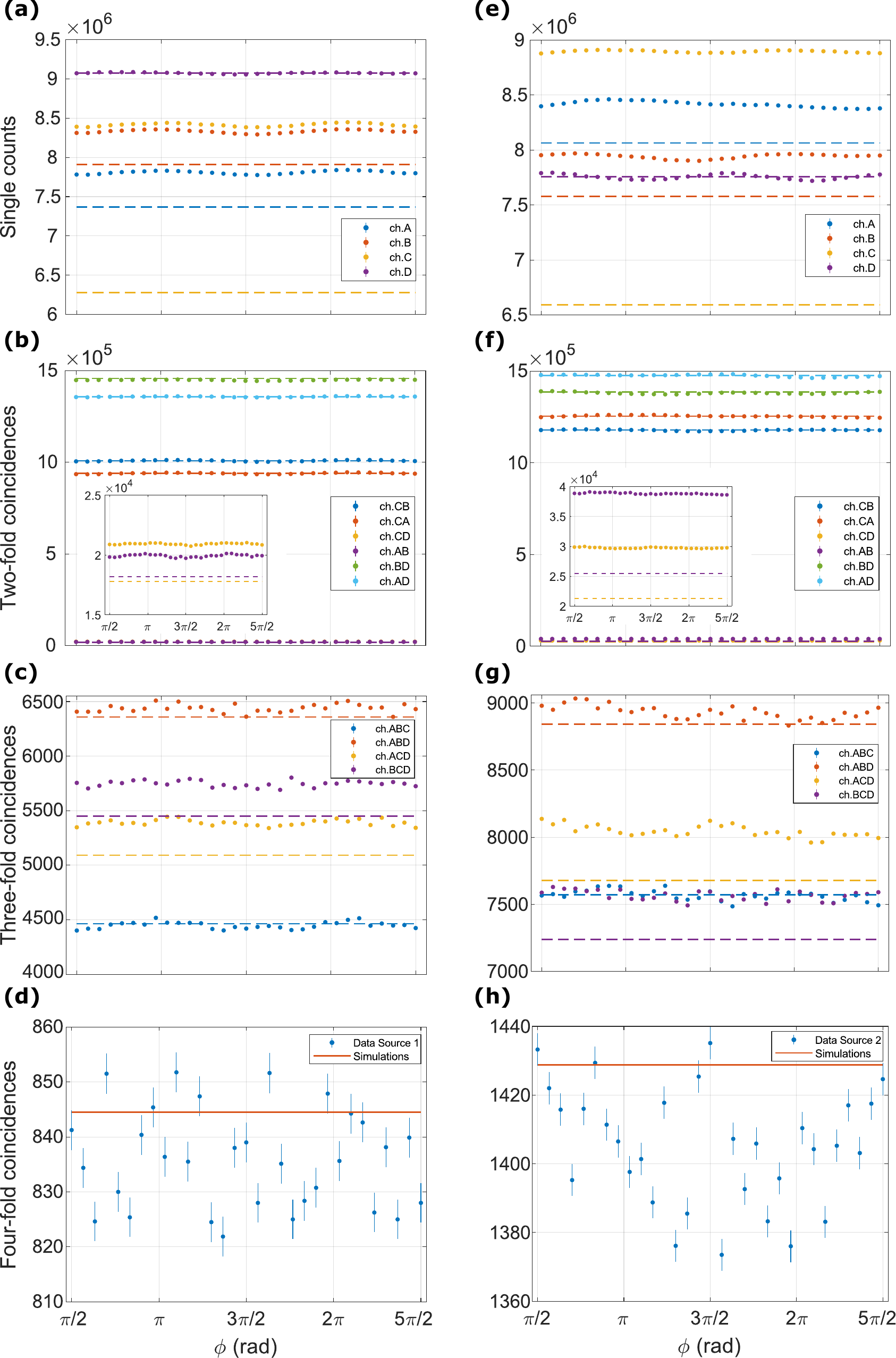}
    \caption{Measurements and simulations of the single-source measurements for $\chi=\pi$ in Eq.~\eqref{eq:inputState}.
    Panels \textbf{(a)}-\textbf{(d)} report single counts and coincidence events measured when blocking the source of separable photons (Source 1).
    The panels also show the corresponding results achieved via simulations based on the optimisation algorithm discussed in Appendix~\ref{appendix:simulations}.    
    Panels \textbf{(e)}-\textbf{(h)} report the same results obtained when blocking the source of entangled photons (Source 2). In all graphs, the integration time for each point is \qty{60}{\second}.}
    \label{fig:measurements and simulations}
\end{figure*}

We present here the raw data of the background-subtracted measurement in Fig.~\ref{fig:background subtraction}.
As described in the main text, we performed $\varphi$-phase scans at fixed phases $\chi=0,\,\pi$, and $\theta=0$ in Eqs.~\eqref{eq:inputState}-\eqref{eq:separable state}.
We first show the results of single-source measurements (when only one source is pumped by the laser) and, subsequently, the results of the experiment when background subtraction is not employed.

Single-source measurements provide a useful comparison for the expected fluctuations of single counts and coincidence events in the absence of interference of photons from different sources.
In addition, since photons from one source are coupled to separate beam splitters (cf. Fig.~\ref{fig:experimental setup}), also no interference effect between photons from the same source can take place, and fluctuations can be purely associated with other contributions, such as residual polarisation dependence of the system.

We report in Fig.~\ref{fig:measurements and simulations} the measurements of single counts and coincidence events (together with the simulations discussed in Appendix~\ref{appendix:simulations}) for the two separate sources when $\chi=\pi$, but similar results are obtained for $\chi=0$.
Moreover, all four-fold background coincidence measurements (for both $\chi=0,\,\pi$) can be found in Figs.~\ref{fig:background subtraction}(a) and \ref{fig:background subtraction}(b).
Single counts in Figs.~\ref{fig:measurements and simulations}(a) and \ref{fig:measurements and simulations}(e) exhibit relative fluctuations of the order of $\simeq 0.2\%$.
Similarly to single counts, two-, three-, and four-fold coincidence events have only marginal fluctuations, respectively of the order of $\simeq 0.3\%$, $\simeq 0.5\%$ and $\simeq 1.2\%$.
The increasing value in the fluctuations of coincidence events is mainly given by the smaller photon counting statistics of these cases.

\begin{figure*}
    \centering
    \includegraphics[width=0.65\textwidth]{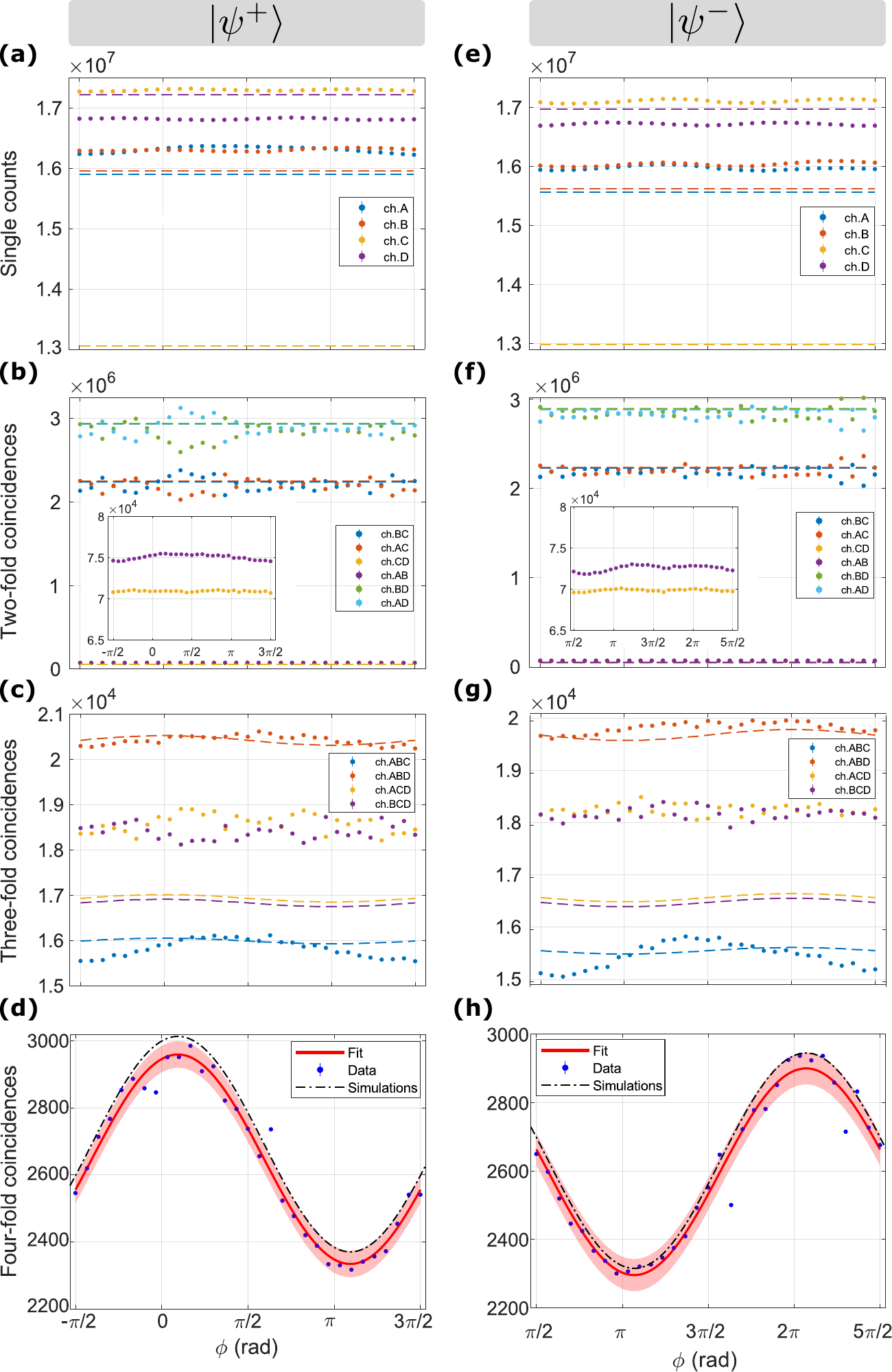}
    \caption{Measurements of single and coincidence counts for two settings of the entangled state phase $\chi=0,\,\pi$ in Eq.~\eqref{eq:inputState}  ($\ket{\psi^+}$ and $\ket{\psi^-}$ in panels (a)-(d) and (e)-(h), respectively). \textbf{(a)(e)} Single counts. \textbf{(b)(f)} Two-fold coincidences. \textbf{(c)(g)} Three-fold coincidences. \textbf{(d)(h)} Four-fold coincidences data (blue dots) fitted with a cosine function (red curve). The red-shaded region shows the fit prediction interval at a confidence level of one standard deviation. The graphs include the results of the multiphoton interference simulations (black dashed curve). The visibility of the fitting cosine is $11.6(5)\%$ and $11.8(4)\%$ for insets (d) and (h), whereas simulations predict a visibility of $12.0(1)\%$ and $11.9(1)\%$, respectively. In all graphs, the integration time for each point is \qty{60}{\second}.}
    \label{fig:main measurements}
\end{figure*}

Fig.~\ref{fig:main measurements} shows the data collected when both sources are employed, before subtraction of the background, for both settings of the phase $\chi=0,\,\pi$ (with $\theta=0$).

The single counts in Figs.~\ref{fig:main measurements}(a) and \ref{fig:main measurements}(e) show only small changes as we vary the collective phase $\phi$, with relative fluctuations of $\simeq 0.1-0.2\%$.
This value is consistent with the corresponding fluctuations of single counts measured when using only one source at a time, where the smaller sample size produces slightly higher values.

Two-fold coincidences in Figs.~\ref{fig:main measurements}(b) and \ref{fig:main measurements}(f) have relative fluctuations of the order of $\simeq 1.7\%$ and $\simeq 2.4\%$, respectively. 
These values are larger than the corresponding ones measured with the single sources ($\simeq 0.3\%$). Nevertheless, they are connected to interference effects between the input states produced by the two sources \cite{Tichy2011} that cannot be attributed to a dependence on the collective phase.
In particular, these fluctuations do not show the expected cosinusoidal dependence on the collective phase, but rather exhibit a complex interference pattern that arises as a consequence of the relative phase between the dominant single-pair emission contributions of the two sources.
In fact, two-fold coincidences of channels associated with the same beam splitter --- chs.~A-B or chs.~C-D in Fig.~\ref{fig:experimental setup} ---, which are not influenced by these contributions, exhibit significantly weaker fluctuations ($\simeq 0.3-0.4\%$), as shown in the insets of Figs.~\ref{fig:main measurements}(b) and \ref{fig:main measurements}(f), despite their reduced photon counting statistic.
Specifically, the partial connectivity of the interferometer does not allow single-pair emissions from one source to contribute to two-fold coincidences of chs.~A-B or chs.~C-D, which necessitate two-pair emissions. 
This requirement prevents two-fold coincidences between these channels from only two particles in the system and ensures contributions from a wider spectrum of interfering terms, originating from all exchange processes of at least four particles at the two beam splitters, thus reducing the influence of relative phase terms and, consequently, possible fluctuations.
Therefore, the coincidences between chs.~A-B and chs.~C-D are a good indicator for the independence of two-fold coincidences from the collective 4P phase.
Of these, chs.~C-D do not exhibit a clear dependence on $\phi$, as shown in the two insets, whereas chs.~A-B have slightly larger fluctuations with a specific pattern displaying a maximum in the centre of the $\phi$ range (corresponding to half the rotation range of the half-wave plate in input mode 1).
In contrast, a pure dependence on the collective phase $\phi$ would lead to a $\pi$-shift, i.e., an inversion of the pattern between the two curves for $\ket{\psi^+}$ and $\ket{\psi^-}$, which is clearly not the case.
Therefore, we conclude that this fluctuation pattern is dominated by artefacts of the wave-plate rotation rather than the collective phase.

Three-fold coincidences in Figs.~\ref{fig:main measurements}(c) and \ref{fig:main measurements}(g) have residual interference effects similar to two-fold coincidences that appear as anti-correlated patterns of coincidences comprising only one of the output channels A and B.
The influence of such interference effects on the three-fold coincidences is already significantly weaker with respect to the two-fold case, with fluctuations of the order of $\simeq 0.8\%$.
Conversely, the largest fluctuations are here connected to three-fold coincidences among chs.~A-B-C, which, however, show a pattern that, in both cases, can be correlated to the two-fold coincidence data of chs.~A-B (Pearson correlation coefficients $\simeq 0.91$).
Consequently, fluctuations are once more associated with unwanted artefacts and residual interference effects, instead of a dependence on the collective phase.

While two-fold and three-fold coincidences do not exhibit any obvious dependence on the collective phase $\phi$, the four-fold coincidences reported in Figs.~\ref{fig:main measurements}(d) and \ref{fig:main measurements}(h) show a clear oscillation.
In particular, the four-fold coincidences are in excellent agreement with the cosine behaviour predicted by Eq.~\eqref{eq:4pointcor_prediction}, showing relatively high visibility compared to the previous cases of $11.6(5)\%$ and $11.8(4)\%$, which excludes the possibility that the observed four-fold oscillations are an artefact of the interference effects mentioned above for two- and three-fold coincidences.
Moreover, as also shown in the main text, the experimental data exhibit a good agreement with numerical simulations of the experiment (further details in Appendix~\ref{appendix:simulations}).

\section{Simulations} \label{appendix:simulations}
To further support the experimental results, we performed simulations of the experiment based on the specific experimental scheme and the photon sources used to witness the entanglement-induced collective interference.

For this purpose, we performed a characterisation of the two SPDC sources and developed a model of the unitary transformation realised in Fig.~\ref{fig:experimental setup}(b).
Source characterisation is required to calculate the emission rates of single-pair emissions and higher-order emissions from the SPDC processes, depending on the laser pump power and pulse repetition rate.
The unitary transformation model, instead, allows us to evaluate the probability of all possible output events for each input state corresponding to combinations of different emission orders from the two sources.
The reconstruction of the unitary transformation is achieved via an optimisation algorithm based on measurements of single count events --- the detection events measured by each detector independently of the measurements of all other detectors --- and all $k$-fold coincidence rates, with $k=2,\, 3,\, 4$, produced by the individual sources.
As a result, we can calculate the total single counts and coincidence rates by using the emission rates of each input state and the relative output probabilities.

The simulations take into account a few fundamental approximations.
First, the two sources are considered independent of each other.
This is strictly true only when we use single counts and coincidence rates from single source emissions, i.e., the data exploited for the reconstruction of the unitary transformation.
In fact, as already mentioned in Sec.~\ref{appendix:raw data}, the presence of phase terms among the input states of the combined two-source SPDC process can give rise to interference in the coincidence rates.
Second, it is assumed that the down-conversion process of the pump photons results in spectrally pure down-converted photon pairs, corresponding to fully separable output photon states --- the production of entangled pairs is only associated with the geometry of the Sagnac interferometers \cite{Fedrizzi2007}.
This approximation is well justified due to the high spectral purity ($\geq 98\%$) of the SPDC sources used in the experiment, as extensively studied in \cite{Pickston2021}.
Partial distinguishability of the input particles (see Eqs.~\eqref{eq:inputState}-\eqref{eq:separable state}) is obtained in terms of polarisation states, which, following the formalism developed in \cite{Dittel2019}, are considered as internal degrees of freedom of the input state.
While this is computationally efficient, the unitary transformation is unable to act on the photons' transformation depending on their polarisation state, thus excluding any modelling of residual polarisation dependence of losses and beam splitters' splitting ratios.
However, this limitation does not significantly influence the results to a first approximation.

\subsection{Photon sources}
The photon sources in Fig.~\ref{fig:experimental setup}(a) are modeled with two independent SPDC processes, where we expressed the resulting photon states in terms of two-mode squeezed vacuum states in the photon number basis $n$:
\begin{equation} \label{eq:two-mode squeezed vacuum}
    \ket{\varphi}=\sqrt{1-\gamma^2}\sum_{n=0}^\infty \gamma^n \ket{n,n}_{s,i},
\end{equation}
with $s$ and $i$ indicating the signal and idler modes, and $\gamma$ the squeezing parameter.
This photon state represents a good approximation for calculating the emission rates even for the bidirectionally pumped crystal in Fig.~\ref{fig:experimental setup}(a) (used to produce entangled pairs).
The emission probability of a specific photon number state $\ket{n,n}_{s,i}$ per laser pulse is associated with the squeezing parameter by $P(n)=(1-\gamma^2)\gamma^{2n}$.
Moreover, the squeezing parameter can be expressed as $\gamma=\sqrt{\tau p}$, in terms of the pump power $p$ and a constant $\tau$ that quantifies the nonlinear interaction within the aKTP crystals \cite{Jin2015,Pickston2021}.
Similarly to \cite{Jin2015}, we calculated $\tau$ for both sources by measuring single counts and coincidence rates and knowing the repetition rate of the laser pulses $f$ ($\qty{80}{\mega\hertz}$). Specifically, by considering an input optical power of $\qty{174}{\milli\watt}$ and $\qty{168}{\milli\watt}$ for Source 1 and Source 2, respectively, we obtained squeezing parameters of $\gamma_{S_1}=0.102$ and $\gamma_{S_2}=0.094$.

The evaluation of $\gamma$ allows us to determine the single photon pair rates and the rates of higher-order pair productions depending on the pump power.
To show this, by using the polarisation basis, the overall state of the down-converted photons collected at the output of the Sagnac interferometers can be written as:
\begin{widetext}
\begin{equation} \label{eq:state squeezing parameter}
    \ket{\psi} = \ket{\varphi_{S_1}} \otimes \ket{\varphi_{S_2}} = \Big( \sqrt{1-\gamma_{S_1}^2} \sum_{n=0}^{\infty} \gamma_{S_1}^n \big[ \frac{1}{\sqrt{2}}(\ket{H,V}_{2,3} + e^{-i\chi}\ket{V,H}_{2,3}) \big] ^{\otimes n} \Big) \otimes \Big( \sqrt{1-\gamma_{S_2}^2} \sum_{m=0}^{\infty} \gamma_{S_2}^m \ket{H,V}_{1,4}^{\otimes m} \Big),
\end{equation}
\end{widetext}
where $S_1$ and $S_2$ refer to the source of entangled photon states (Source 1) and of separable photon states (Source 2), the indices $1,~2,~3,~4$ correspond to chs.~1-4 before the beam splitters in Fig.~\ref{fig:experimental setup}(b), and $\chi$ is the usual phase of the entangled state as defined in Eq.~\eqref{eq:inputState}.
Starting from Eq.~\eqref{eq:state squeezing parameter}, we can calculate the emission rates of each mode-occupation list $\mathbf{R}=(r_1,r_2,r_3,r_4)$ corresponding to the occupation number of each unitary input channel $1-4$ \cite{Dittel2019}.
Given the nature of the SPDC process (two-photon emissions) and the arrangement of the channels in Fig.~\ref{fig:experimental setup}, we have $r_1=r_4:=r_{S_2}$ and $r_2=r_3:=r_{S_1}$, where $r_{S_1}$ and $r_{S_2}$ are defined as the number of pair emissions of each source.
Therefore, each mode-occupation list $\mathbf{R}$ can be associated with an input state similar to Eq.~\eqref{eq:inputState}:
\begin{equation}
    \begin{split}\label{eq:inputState simulations}
    &\ket{\xi} = \left[ \frac{1}{\sqrt{2}} ( \hat{a}_{2,H}^{\dag} \hat{a}_{3,V}^{\dag} + e^{-i\chi} \hat{a}_{2,V}^{\dag} \hat{a}_{3,H}^{\dag}) \right]^{r_{S_1}}\cdot\\
    &\qquad \quad (\hat{a}_{1,\mathrm{S}}^{\dag}(\varphi) \hat{a}_{4,\mathrm{S}}^{\dag}(\theta))^{r_{S_2}} \ket{0}.
    \end{split}
\end{equation}
The state $\ket{\xi}$ is related to the contribution of a specific pair of values $(m,n)$ in Eq.~\eqref{eq:state squeezing parameter}, from which one can evaluate the emission rate by knowing $\gamma_{S_1}$, $\gamma_{S_2}$, and $f$.

For example, the input state in Eq.~\eqref{eq:inputState} is associated with $r_{S_1}=r_{S_2}=1$, $\mathbf{R}=(1,1,1,1)$, i.e., a single firing of both sources.
The contribution to this state is related to the $m=n=1$ terms in Eq.~\eqref{eq:state squeezing parameter}, whose emission rate is given by $r_{1,1} = f P(\mathbf{R}_{1,1}) = f (1-\gamma_{S_1}^2)\gamma_{S_1}^2 (1-\gamma_{S_2}^2)\gamma_{S_2}^2$.
In a similar fashion, one can calculate the rates of all mode-occupation lists resulting from all possible combinations of sources' emissions.
We considered here contributions of up to three photon-pair productions, which can arise from both single-source emissions ($r_{S_1}=3,~r_{S_2}=0$, or the opposite) or emissions from combinations of both sources ($r_{S_1}=2,~r_{S_2}=1$, or the opposite).

\subsection{Unitary reconstruction}
Each mode-occupation list contributes differently to the total single counts and coincidence rates.
In fact, each mode-occupation list has a different emission rate and probability of giving rise to specific output events after propagation through the interference setup.
The calculation of these probabilities is based on the formalism developed in \cite{Dittel2019}, which, given a certain unitary transformation, provides the input-output probabilities for each input mode-occupation list $\mathbf{R}$.

We reconstructed the unitary transformation associated with Fig.~\ref{fig:experimental setup}(b), given by the two independent beam splitters, to include the possibility of both optical losses before and after the two beam splitters, and unbalanced splitting ratios of the beam splitters themselves.
Losses are modelled by extending the four-mode unitary with additional unbalanced beam splitters and external modes.
We consider independent transmission factors $\eta_j$ for each of the eight input and output modes of the beam splitters.

The multiplexed eight-detector scheme (inset of Fig.~\ref{fig:experimental setup}(b)) is modelled via additional balanced beam splitters, randomly distributing the arriving photons on the two detectors of each output channel A-D. This is a valid simplification, as in the experiment the polarisation projection is only used to fix the photon polarisation on individual detectors. All losses of this second transformation stage and the detector efficiencies are incorporated into the output channel losses of the main beam splitter unitary.

As a result, ten parameters are required to reconstruct the extended unitary: eight transmission coefficients $\eta_j$ to model the input and output losses and the two splitting ratios of the main beam splitters.
We employed an optimisation algorithm for these ten parameters, which aims to match the average values of the simulated and measured single counts and coincidence rates.
More precisely, we carried out this parameter search by simultaneously emulating both single-source measurements performed by blocking one source at a time --- thus matching eight single count rates, twelve two-fold coincidence rates, eight three-fold coincidence rates, and two four-fold coincidence rates.
This procedure is used for both the two measurements with $\chi=0,\,\pi$ to allow for marginal deviations in alignment between the two experimental conditions.

The panels in Fig.~\ref{fig:measurements and simulations} show the comparison between measurements and simulations for the case $\chi=\pi$, and similar results are obtained for $\chi=0$. 
Consistently, the transmission coefficients of the main beam splitters do not differ between the two optimisations to all reasonable orders of approximation, and variations of input and output losses are of the order of $\pm 0.8\%$.
Moreover, for the specific beam splitters used, we found deviations from a balanced 50:50 splitting ratio that are within the $\pm 3\%$ tolerance interval given by the producer \cite{NewportBS}.
The panels associated with the coincidence rates in Fig.~\ref{fig:measurements and simulations} show good agreement with the measurements, with relative deviations $<5\%$.
This is only partially true for single count rates, where channel C has a relative deviation of $\simeq 25\%$.
Presumably, this is an effect of correlated loss affecting channel C, e.g. due to small errors in setting the FPGA unit's logic delays, which is hardly noticeable beforehand without applying this optimisation routine.
This defect is modelled by the algorithm with an effective higher loss of channel C.

\subsection{Power drifts} \label{appendix:power drift}
As mentioned in Sec.~\ref{sec:experiment}, drifts in laser power between the main measurements in Fig.~\ref{fig:main measurements} and the background measurements in Fig.~\ref{fig:measurements and simulations} lead to errors in the background subtraction.

Therefore, we performed another optimisation routine based on the single counts and all $k$-fold coincidence rates of the main measurements (again one optimisation for each phase $\chi=0,\,\pi$ in Eq.~\eqref{eq:inputState}), but this time only allowing for a variation of the input power affecting both photon sources in the same manner.
According to this procedure, the power is approximately unchanged for the measurement of $\chi=\pi$, demonstrating that no significant power fluctuation influences Fig.~\ref{fig:background subtraction}(d), while a relative power drift of $\simeq2.3\%$ affects the measurement of $\chi=0$, thus resulting in the considerable visibility reduction obtained in Fig.~\ref{fig:background subtraction}(c).

\end{document}